\documentstyle[preprint,tighten,version2,aps]{revtex}
\def\goto{\mathop{\;\longrightarrow\;}}
{\makeatletter%
\gdef\labeleqs#1{{%
\edef\@currentlabel{%
\ifappendixon\appletter\fi
\ifsecnumbers\ifnum\c@secnum>0
\arabic{secnum}.\fi\fi\arabic{equation}}%
\label{#1}%
}}%
}
\begin{document}
\draft
\preprint{IFUP-TH 8/94}
\begin{title}
Two dimensional $SU(N)\times SU(N)$ Chiral Models on the Lattice (II):\\
the Green's Function.
\end{title}
\author{Paolo Rossi and Ettore Vicari}
\begin{instit}
Dipartimento di Fisica dell'Universit\`a and
I.N.F.N., I-56126 Pisa, Italy
\end{instit}
\begin{abstract}
Analytical and numerical methods are applied to principal chiral models on
a two-dimensional lattice and their predictions are tested and compared.
New techniques for the strong coupling expansion of $SU(N)$ models are
developed and applied to the evaluation of the two-point correlation
function. The momentum-space lattice propagator is constructed with
precision $O\left( \beta^{10}\right)$ and an evaluation of the correlation
length is obtained for several different definitions.

Three-loop weak coupling contributions to the internal energy and to
the lattice $\beta$ and $\gamma$ functions are evaluated for all $N$, and
the effect of adopting the ``energy'' definition of temperature is
computed with the same precision. Renormalization-group improved
predictions for the two-point Green's function in the weak coupling
(~continuum~) regime are obtained and successfully compared with Monte
Carlo data.

We find that strong coupling is predictive up to a point where asymptotic
scaling in the energy scheme is observed. Continuum physics is insensitive
to the effects of the large $N$ phase transition occurring in the lattice
model.
Universality in $N$ is already well established for $N\gtrsim 10$
and the large $N$ physics is well described by a ``hadronization''
picture.

\end{abstract}


\narrowtext

\section{Introduction}
\label{intro}

The study of two dimensional $SU(N)\times SU(N)$ principal chiral models
is strongly motivated by the deep analogies between this class of field
theories and four dimensional non Abelian gauge theories.
We only mention here asymptotic freedom and the existence of a large $N$
limit that can be represented as a sum over planar diagrams.
A standard lattice version of the continuum action
\begin{equation}
S\;=\;\int d^2x\,{1\over T}\,{\rm Tr}\,\partial_\mu U(x)\partial_\mu
U^\dagger (x)
\label{caction}
\end{equation}
is obtained by introducing a nearest-neighbour interaction
\begin{equation}
S_L\;=\;-2N\beta\,\sum_{n,\mu} {\rm Re}\,{\rm Tr}\,\left[ U_n
U^\dagger_{n+\mu}\right]\;,\;\;\;\;\;\beta\,=\,{1\over NT}\;\;\;.
\label{laction}
\end{equation}
No exact solution of these models is known, even in the large $N$ limit.
An exact S-matrix has however been conjectured
\cite{Abdalla,Wiegmann1,Wiegmann2}, and numerical evidence seems to
indicate that the corresponding bound state spectrum is reproduced in the
continuum limit \cite{chiral1}.
Also the mass-$\Lambda$ parameter ratio has been conjectured by using the
Bethe Ansatz approach, and the result is \cite{Balog}
\begin{equation}
R_{\overline{MS}}\;\equiv\;{M\over\Lambda_{\overline
{MS}}}\;=\;\sqrt{{8\pi\over e}} \;
{\sin \pi/N\over \pi/N}\;\;\;.
\label{mass-lambda}
\end{equation}

Information coming from S-matrix and large $N$ factorization leads to the
conclusion that when $N\rightarrow\infty$
principal chiral models are just free field theory in disguise.
In other words, a local nonlinear mapping should exist between the
Lagrangian fields $U$ and some Gaussian variables
\cite{Polyakov}. The nontriviality of the realization may however be
appreciated when considering the two-point Green's function of the
Lagrangian field: while at small Euclidean momenta there is substantial
evidence for an essentially Gaussian (free field) behavior, at large
momenta, where results from standard weak coupling
perturbation theory are expected to hold by asymptotic freedom,
the Lagrangian field seems to
behave more like a composite object formed by two elementary (Gaussian)
excitations, which cannot however appear like deconfined free particles.
This elementary ``hadronization'' picture, where however the Lagrangian
fields themselves play the role of hadrons, is strongly supported by all
numerical evidence that we have produced, and for $N\gtrsim 6$ is universal,
i.e. independent of $N$, which confirms that the $1/N$ expansion, were it
available, would be an extremely predictive tool in the analysis of these
models.

In the persistent absence of such an expansion, we here tried to apply all
other analytical and numerical methods of lattice field theory that seemed
appropriate to the problem at hand.
In particular we sistematically extended the strong coupling
character expansion to $SU(N)$ groups, generating compact formulas for many
``standard'' integrals and character coefficients and extending the series
for the free and internal energy, the magnetic susceptibility and various
definitions of the mass gap.
Since principal chiral models possess a wide scaling window in the strong
coupling domain, these results become a powerful tool for the analytic
computation of physical (continuum) quantities with very small systematic
error due to scaling violations.

We also performed many three-loop weak coupling lattice computations in
order to improve our understanding of the approach to asymptotic scaling.
We found that introducing a new coupling $T_E$ proportional to the energy
\cite{Parisi},
the so called ``energy''
variable, allows an amazingly impressive improvement
in the convergence of lattice results towards the continuum
``asymptotic scaling'' predictions, not only when considering such ratios
as $M/\Lambda_L$, but also when parametrizing the running coupling
dependence of the two-point Green's function in the large momentum regime.

Some of the numerical and analytical results presented here were announced
(and presented without proof) in our Refs.~\cite{chiral1,chiral1c}.
The study of the two-point Green's function in turn is fully original and
constitutes another impressive test of scaling and universality in the
context of large $N$ principal chiral models.

This paper is organized as follows.

In Section~\ref{strongcoupling} we describe a new technique for the
strong coupling expansion of $SU(N)$ models. The strong coupling
series of the free-energy density is calculated up to $O\left(
\beta^{14}\right)$ for $N>7$.

In Section~\ref{strongcoupling2} we apply our new technique to
the evaluation of the two-point correlation function. Strong coupling
series of several different definitions of correlation length are presented.

In Section~\ref{weakcoupling} we compute the three-loop weak coupling
contributions to the internal energy, and to the lattice $\beta$-function
and anomalous dimension of the fundamental field.
The corresponding quantities in the energy scheme are also calculated.
Continuum predictions for the two-point
correlation function are obtained by solving the corresponding renormalization
group
equation.

In Section~\ref{numericalresults} we present the results of Monte Carlo
simulations for several large values of $N$, and compare them to our
analytical (strong and weak coupling) calculations.

\section{Strong coupling expansion of $SU(N)$ models}
\label{strongcoupling}

As stated in the introduction, a renewed interest in the strong coupling
expansion of chiral models was stimulated by the observation of precocius
scaling well within the expected
convergence radius of the strong coupling series \cite{chiral1},
and by the relevance of the complex $\beta$-singularities of the partition
function close to the real axis \cite{Marinari}.

Strong coupling in matrix-valued lattice models was pioneered
many years ago by several authors. Most studies were however addressed to
the (relatively simpler) problem of computing $U(N)$ group integrals
\cite{BarsGreen,Samuel,Bars,Brower}, while not many
results are available concerning $SU(N)$ integration
\cite{Creutz,ShigKogut,Brower2}.
Moreover, while very general compact formulas can sometimes be written,
often these formulas require a lot of supplementary work in order to
extract the directly relevant information.
Alternatively, tables of numerical coefficients can be routinely generated
by computer programs \cite{Hansen}, but generality of the results is
completely lost. In the search for sufficiently general, but at the same
time manageable results we tried to follow a pathway originally opened
in Refs. \cite{Green,Rossi,Guha}, whose notation we shall try to follow as
far as possible.

We shall focus on the free-energy density of the one
dimensional $SU(N)\times SU(N)$ chiral model, which can be reinterpreted
as a generating functional for $SU(N)$ group integral in the ``standard
form''
\begin{equation}
S_{p,q}\;=\; \int dU \; \left( {\rm Tr} \,U \right)^p
\,\left( {\rm Tr} \,U^\dagger \right)^q \;\;,
\label{s1}
\end{equation}
where $dU$ is the normalized Haar measure for $SU(N)$.
According to the definitions, the free-energy density $F$ can be
obtained from evaluating
\begin{equation}
\exp F(\beta) \;=\;\int dU\; \exp \left[ N\beta \left(
{\rm Tr} \,U \,+\, {\rm Tr} \,U^\dagger \right) \right]\;\;.
\label{s2}
\end{equation}
In turn knowledge of $F$ allows the knowledge of the coefficients
$z_{(r)}(\beta)$ of the character expansion of the integrand:
\begin{equation}
\exp \left[ N\beta \left(
{\rm Tr} \,U \,+\, {\rm Tr} \,U^\dagger \right) \right]\;=\;
\exp \left[ - F(\beta) \right] \sum_{(r)} d_{(r)}
z_{(r)}\chi_{(r)} (U) \;\;,
\label{s3}
\end{equation}
 where $\sum_{(r)}$ is a sum over all finite dimensional irreducible
representations of the group, $\chi_{(r)}$ and $d_{(r)}$ are the
corresponding characters and dimensions.
Let's now recall from Ref. \cite{Rossi} a few
exact results concerning the $U(N)$ integration.
If we denote by
\begin{equation}
A_{m,N} \;\equiv\; {\rm det} \left[ I_{m+i-j} (s) \right]\;\;,
\label{s4}
\end{equation}
where $I_l$ are the modified Bessel functions and $s=2N\beta$, we find that
\begin{equation}
\widetilde{F}_N(\beta)\;=\; \ln A_{0,N}(s)\;\;,
\label{s5}
\end{equation}
\begin{equation}
\Delta_{m,N}(\beta)\;\equiv\; \langle \left[ {\rm det}\,U \right]^m \rangle
\;=\; {A_{m,N}\over A_{0,N}}\;\;,
\label{s6}
\end{equation}
and the following non linear ordinary differential equations are satisfied:
\begin{equation}
{1\over s}{d\over ds}s{d\over ds}\Delta_{1,N}\,+\,
{1\over 1-\Delta_{1,N}^2} \left[ \left( {d\Delta_{1,N}\over ds}\right)^2
-{N^2\over s^2}\right] \Delta_{1,N}\,+\, \left(1-\Delta_{1,N}^2\right)
\Delta_{1,N}\;=\;0\;\;,
\label{equation}
\end{equation}
\begin{equation}
{d\over ds} \left( \ln F_N - \ln F_{N-1}\right)\;=\;
{\Delta_{1,N}\over 1-\Delta_{1,N}^2 }\left( {d\Delta_{1,N}\over ds} +
{N\over s}\Delta_{1,N}\right)\;\;.
\label{s8}
\end{equation}
Moreover \cite{Guha}
\begin{equation}
\Delta_{m,N}^2\,-\, \Delta_{m+1,N} \Delta_{m-1,N}\;=\;
\Delta_{m,N-1}\Delta_{m,N+1}\left( 1-\Delta_{1,N}^2 \right)\;\;.
\label{s9}
\end{equation}
As a consequence of these equations, one may obtain the following
strong coupling $U(N)$ results \cite{Rossi,Guha}:
\begin{equation}
\Delta_{1,N}(\beta)\;=\; J_N(2N\beta) \,+\,O\left( \beta^{3N+2}\right)\;\;,
\label{s10}
\end{equation}
\begin{equation}
\Delta_{2,N}(\beta)\;=\; J_N(2N\beta)^2 \,-\, J_{N-1}(2N\beta)J_{N+1}(2N\beta)
 \,+\,O\left( \beta^{4N}\right)\;\;,
\label{s11}
\end{equation}
\begin{equation}
\widetilde{F}_N(\beta) \;=\;N^2\beta^2\,-\, \sum_{k=1}^\infty
kJ_{N+k}(2N\beta)^2 \,+\,O\left( \beta^{4N+4}\right)\;\;.
\label{s12}
\end{equation}
The passage to $SU(N)$ is effected by the relationship
\begin{equation}
F_N(\beta) \;=\; \widetilde{F}_N(\beta)\,+\,
\ln \sum_{k=-\infty}^\infty \langle \;\left( {\rm det}\, U \right)^k \;\rangle
\;\;,
\label{s13}
\end{equation}
implying
\begin{equation}
F_N(\beta) \;=\; N^2\beta^2\,+\, 2J_N(2N\beta)\,-\,
2J_{N-1}(2N\beta) J_{N+1}(2N\beta) \,-\,
\sum_{k=1}^\infty k J_{N+k}^2 (2N\beta)\,+\,O\left( \beta^{3N}\right)\;\;.
\label{s14}
\end{equation}
We can now make use of known relationships concerning the series
expansion of a product of Bessel functions to obtain a wide number of
``standard'' integrals:
\begin{equation}
S_{p,p}\;=\;p!\;\;,\;\;\;\;\;p\leq N\;\;;
\label{s15}
\end{equation}
\begin{equation}
S_{p,N+p}\;=\;\sum_{q=0}^p (-1)^{q}
 {(N+p)!\over (N+q)!}
\left( \begin{array}{c} p\\ q\end{array}\right)\;\;,\;\;\;\;\; p\leq N+1\;\;;
\label{s17}
\end{equation}
\begin{equation}
S_{N+p,N+p}\;=\;(N+p)!\,+\,\sum_{q=1}^p {(-1)^{q}\over (p-q)!}
\left[ {(N+p)!\over (N+q)!}\right]^2
\left( \begin{array}{c} 2N+2q-2\\ q-1\end{array}\right)\;\;,\;\;\;\;\; p\leq
N+1\;\;;
\label{s16}
\end{equation}
\begin{equation}
S_{p,2N+p}\;=\;\sum_{q=0}^p (-1)^{q}
 {(2N+2q)! (2N+p)!\over (2N+q)! (N+q+1)! (N+q)!}
\left( \begin{array}{c} p\\ q\end{array}\right)\;\;,\;\;\;\;\; p\leq N+1\;\;.
\label{s18}
\end{equation}
One may check that essentially all Hansen's results \cite{Hansen}
are correctly reproduced and extended to arbitrary $N$
by the above formulas.
Eqs. (\ref{s15}-\ref{s18}) are our first set of compact results
concerning
$SU(N)$ integration.

Even more important is however the possibility of extracting similar
results for the
coefficients of the character expansion. Let's briefly recall from
Ref. \cite{Green} the following expression of $U(N)$ character coefficients
$\widetilde{z}_{(r)}(\beta)$
\begin{equation}
d_{\lambda_1...\lambda_N}\,\widetilde{z}_{\lambda_1...\lambda_N}\;=\;
{{\rm det} \left[ I_{\lambda_i+j-i}(2N\beta) \right] \over
{\rm det} \left[ I_{j-i}(2N\beta) \right]}\;\equiv\; \langle\;
\chi^\ast_{(r)}(U)\;\rangle \;\;.
\label{s19}
\end{equation}
Its $SU(N)$ counterpart can be represented in the form
\begin{equation}
z_{(r)}\;=\; { \sum_{s=-\infty}^\infty \widetilde{z}_{(r+s\cdot 1^N)}
\over \sum_{s=-\infty}^\infty \widetilde{z}_{(s\cdot 1^N)} }\;\;,
\label{s20}
\end{equation}
where by definition
\begin{equation}
\widetilde{z}_{(s\cdot 1^N)}\;=\; \langle\; \left( {\rm det}\,
U\right)^s\;\rangle\;\;,
\label{s21}
\end{equation}
\begin{equation}
d_{(r)}
\widetilde{z}_{(r+s\cdot 1^N)}\;=\; \langle\; \left( {\rm det}\, U\right)^s
\chi^\ast_{(r)}(U)\; \rangle\;\;,
\label{s22}
\end{equation}
and all quantities are computed with the $U(N)$ measure.
These results become however rapidly useless with growing $N$,
due to the intractability of the determinants. We may however
easily recover from the previous results the coefficient of the fundamental
character:
\begin{eqnarray}
&&Nz_1(\beta)\;=\; {\partial\over \partial (2N\beta)}F_N(\beta) \nonumber \\
&&=\;N\beta\,+\, J_{N-1}(2N\beta)\,-\,J_{N+1}(2N\beta)\,-\,
\sum_{k=0}^\infty  J_{N+k}(2N\beta)J_{N+k+1}(2N\beta)\,+\,O\left(
\beta^{3N-1}\right)\;\;.
\label{s23}
\end{eqnarray}
Moreover we may want restrict our attention to the first $\sim 2N$ orders
of the strong coupling series, in order to exploit the techniques discussed
in the first part of this section.
In this case the character coefficients of $U(N)$ can be reconstructed by
the use of Schwinger-Dyson equations \cite{Green}.
Appendix~\ref{AA} is devoted to a presentation of our main results concerning
the compact evaluation of $U(N)$ and $SU(N)$  character coefficients.

For what concerns the physical problem of evaluating the d-dimensional
free-energy density in $SU(N)\times SU(N)$ chiral models within the strong
coupling expansion, we address the reader to Ref.~\cite{Green}
for a discussion of the relationship between the $SU(N)$ and the $U(N)$
character expansions. Suffice it to say that knowing the $U(N)$
expansion to $O\left( \beta^{2m}\right)$ allows an immediate identification
with the corresponding $SU(N)$ expansion when $N > m$.
When $N=m$ only a minor modification is required in order to avoid a
double-counting due to the self-duality of the representation
$(1^{N/2};0)=(0;1^{N/2})$. When $N < m$ the procedure must be much more
careful, but in any case this condition would violate the already mentioned
restriction on the applicability of our approach, which does not trivially
extend to more than $2N$ orders of the strong coupling series.

In order to exhibit some new results concerning the strong coupling series
for the free-energy of the principal chiral models, let's first introduce
a notion of ``potential'' intimately related to that presented by Green and
Samuel in Ref.~\cite{Green}
\begin{equation}
W_{st}\;=\; \left[
z^s_{(2;0)} d^t_{(2;0)}\,+\,z^s_{(1^2;0)}d^t_{(1^2;0)}\,+\,
2^{1-t}z^s_{(1;1)}d^t_{(1;1)}\,-\,2^{2-t}N^{2t}z^{2s}_{1}\right]N^{2-2t}\;\;.
\label{t1}
\end{equation}
For comparison, we mention that
\begin{equation}
W_{st}\;=\; V_{st} \,+\, 2^{1-t}\bar{V}_{st}\;\;,
\label{t2}
\end{equation}
where $V$ and $\bar{V}$ have been defined in Ref.~\cite{Green}.
The tenth order strong coupling character expansion for the free-energy in
d-dimensions is then \cite{Barsanti}
\begin{eqnarray}
{\cal F}\;&&=\;
\left( \begin{array}{c} d\\ 1\end{array}\right){F_N\over N^2}\,+\,
2\left( \begin{array}{c} d\\ 2\end{array}\right)\left(z_1^4+W_{42}\right)\,+\,
\left[ 4\left( \begin{array}{c} d\\ 2\end{array}\right)+32
\left( \begin{array}{c} d\\ 3\end{array}\right)\right]z_1^6\,+\nonumber \\
&&\left[ 8\left( \begin{array}{c} d\\ 2\end{array}\right)+48
\left( \begin{array}{c} d\\ 3\end{array}\right)\right]
\left[ {1\over 2}W_{11}\,z_1^6 + W_{31}\,z_1^4\right]\,+\,
\left[ 14\left( \begin{array}{c} d\\ 2\end{array}\right)+372
\left( \begin{array}{c} d\\ 3\end{array}\right)+1296
\left( \begin{array}{c} d\\ 4\end{array}\right)
\right]z_1^8\,+\nonumber \\
&&\;\;96\left( \begin{array}{c} d\\ 3\end{array}\right)W_{21}\,z_1^6\,+\,
\left[ 24\left( \begin{array}{c} d\\ 2\end{array}\right)+576
\left( \begin{array}{c} d\\ 3\end{array}\right)+1536
\left( \begin{array}{c} d\\ 4\end{array}\right)
\right]W_{11}\,z_1^8\,+\nonumber \\
&&\left[ 56\left( \begin{array}{c} d\\ 2\end{array}\right)+4656
\left( \begin{array}{c} d\\ 3\end{array}\right)+46272
\left( \begin{array}{c} d\\ 4\end{array}\right)+95232
\left( \begin{array}{c} d\\ 6\end{array}\right)
\right]z_1^{10}\;\;.
\label{t3}
\end{eqnarray}
Eq.~(\ref{t3}) holds for all $U(N)$ and for $SU(N)$ when $N\ge 5$.
The $SU(N)$ result for the ``potential'' $W_{st}$ can be obtained in the
form
\begin{eqnarray}
&&W_{st}\;\simeq\;
{\beta^{2s}\over 2^t} \left[1+{4s\over N\beta}J^\prime_N(2N\beta)\right]
N^2\left[ \left( 1+{1\over N}\right)^{t-s}+\left( 1-{1\over N}\right)^{t-s}+
2\left( 1-{1\over N^2}\right)^{t-s}-4\right]\,+\nonumber \\
&&\beta^{2s-2}{2s\over 2^t}N^2\left[
J_{N+2}(2N\beta)\left( 1+{1\over N}\right)^{t-s}+
J_{N-2}(2N\beta)\left( 1-{1\over N}\right)^{t-s}-
2J_{N}(2N\beta)\left( 1-{1\over N^2}\right)^{t-s}\right].
\label{t5}
\end{eqnarray}
By explicitly expanding in powers of $\beta$ the $14^{th}$ order character
expansion for the free energy (presented in Ref.~\cite{Green} up to
$O\left( \beta^{12}\right)\,$),
we obtain for $U(N)$ models ($N\ge 7$)
\begin{eqnarray}
&&\widetilde{\cal F}\;=\;
2\beta^2 \, + \, 2\beta^4 \, + \, 4\beta^6
\,+\, \Biggl[ 14+{N^2(5N^2-2)\over (N^2-1)^2 }\Biggr] \beta^8 \, + \, \Biggl[56
+{8N^2 (5N^2-2)\over (N^2-1)^2 }\Biggr]\beta^{10} \nonumber \\
&&+ \Biggl[ 248+
{8N^2(35N^2-17)\over (N^2-1)^2 } + {2N^2(14N^6-11N^4+8N^2-2)\over (N^2-1)^4}
+{16N^4(9N^4-26N^2+8)\over 3(N^2-1)^2 (N^2-4)^2}\Biggr] \beta^{12}
\nonumber \\
&&+\Biggl[ 1176 + {432N^2\over N^2-1} - {32N^4\over (N^2-1)^2}
+{240N^2(5N^2-2)\over (N^2-1)^2}+ {N^2(248N^4-144N^2+48)\over (N^2-1)^3}
\nonumber \\
&&\;\;\;\; +{N^2(436N^6-344N^4+208N^2-48)\over (N^2-1)^4}
+{64N^4(9N^4-26N^2+8)\over (N^2-1)^2(N^2-4)^2}\Biggr]\beta^{14}
+\; O\left( \beta^{16}\right)\;,
\label{t6}
\end{eqnarray}
and for the $SU(N)$ models ($N\ge 7$)
\begin{eqnarray}
&&{\cal F}\;=\;\widetilde{\cal F}\;+\;
{4N^{N-2}\over N!}\beta^N \,+\, \left[ -
{4N^N\over (N+1)!}+{8N^{N-1}\over N!}\right]\beta^{N+2} \;+\nonumber \\
&&\left[ {2N^{N+2}\over(N+2)!}
- {8(N+2)N^N\over (N+1)!} + {4N^{N+1}\over (N-1)N!} +
{24N^{N-1}\over N!} + {4N^{N-2}\over (N-2)!}\right]\beta^{N+4} \;+\nonumber \\
&&\Biggl[ - {2N^{N+4}\over 3(N+3)!} + {4N^{N+2}(N+4)\over (N+2)!} -
{4N^{N+1}(N^2+4N+2)\over (N+1)(N+1)!} + {8N^{N+1}(1+N-N^2)\over (N^2-1)(N+1)!}
  \nonumber \\
&&\;\;+{4N^{N+3}\over 3(N-1)(N-2)N!}
-{24N^{N}(N+2)\over (N+1)!} - {4N^{N}\over (N-1)!} - {8N^{N}\over N!} +
{24N^{N+1}\over (N-1)N!}   \nonumber \\
&&\;\;+ {4N^{N}\over (N-1)^2(N-3)!}
+{112N^{N-1}\over N!} + {24N^{N-2}\over (N-2)!} \Biggr] \beta^{N+6}\;+\;
O\left( \beta^{N+8}\right)\;.
\label{t7}
\end{eqnarray}
The $SU(6)$ result, including an analysis of the $O\left( \beta^{2N}\right)$
contribution, is finally
\begin{equation}
{\cal F}\;=\; 2\,\beta^2 \,+\, 2\,\beta^4 \,+\, {56\over 5}\,\beta^6\,+\,
{84038\over 1225} \,\beta^8 \,+\, {459308\over 1225}\,\beta^{10} \,+\,
{548436429\over 85750}\,\beta^{12}\,+\,...
\label{t8}
\end{equation}
The strong coupling series of the internal energy $E$ can be obtained by
\begin{equation}
E\;=\; 1\;-\;{1\over 4} {d{\cal F}\over d\beta}\;\;.
\label{t9}
\end{equation}

In order to extend the strong coupling series of thermodynamical functions
to higher orders and in order to compute the strong coupling series of
correlation
functions, it will prove convenient to define more involved potentials
corresponding to nontrivial loop topologies and three-body interactions.
For the purposes of the present paper, and in view of further developments,
we shall define
\begin{equation}
\widetilde{W}_{ab}\;=\;z_{(1;1)}^{a+b}\,+\,
N^2\left[ z_{(1;1)}^a{d_{(1;1)}\over 4} \left( {z_{(2;0)}^b\over d_{(2;0)}}
+{z_{(1^2;0)}^b\over d_{(1^2;0)}}\right)
+z_{(1;1)}^b{d_{(1;1)}\over 4} \left( {z_{(2;0)}^a\over d_{(2;0)}}
+{z_{(1^2;0)}^a\over d_{(1^2;0)}}\right)-2z_1^{2a+2b}\right]\;,
\label{t10}
\end{equation}
\begin{eqnarray}
W_{abc}\;=&&
z_{(1;1)}^a\left( z_{(2;0)}^b d_{(2;0)}+z_{(1^2;0)}^b d_{(1^2;0)}\right)
\left( z_{(2;0)}^c d_{(2;0)}+z_{(1^2;0)}^c d_{(1^2;0)}\right)
\;-\;2N^2W_{a1}z_1^{2b+2c}
\nonumber \\
&&-\,z_{(1;1)}^a\left( z_{(2;0)}^{b+c} d_{(2;0)}+z_{(1^2;0)}^{b+c}
d_{(1^2;0)}\right)
\;+\;\;{\rm permutations}\;\;{\rm of}\;\;\;(a,b,c)\nonumber \\
&& + \,z_{(1;1)}^{a+b+c}\left( d_{(1,1)}^2 -d_{(1;1)}\right)
\,-\,4N^4z_1^{2a+2b+2c}\;\;,
\label{t11}
\end{eqnarray}
and the three-body potentials
\begin{eqnarray}
V_{qst}\;=\;N^{1-2t}\;\Biggl[&&
\left(
z_{(3;0)}^qd_{(3;0)}+z_{(2,1;0)}^qd_{(2,1;0)}+z_{(2,1)}^qd_{(2;1)}\right)
z_{(2;0)}^sd_{(2;0)}^t\nonumber \\
&&+\left(
z_{(2,1;0)}^qd_{(2,1;0)}+z_{(1^3;0)}^qd_{(1^3;0)}+z_{(1^2;1)}^qd_{(1^2;1)}\right)
z_{(1^2;0)}^sd_{(1^2;0)}^t
\nonumber \\
&&+2^{1-t}\left( z_{(2;1)}^qd_{(2;1)}+z_{(1^2;1)}^qd_{(1^2;1)}\right)
z_{(1;1)}^sd_{(1;1)}^t
-2N^{2t+1}z_1^qW_{q+s,t+1}\nonumber \\
&&-2^{2-t}N^{2t+1}z_1^{q+2s}W_{q,1}-2^{2-t}
N^{2t+3}z_1^{3q+2s}\Biggr]\;.
\label{t12}
\end{eqnarray}

\section{The strong coupling expansion of correlation functions}
\label{strongcoupling2}

A typical application of the strong coupling analysis amounts to the
evaluation of the so-called ``true mass gap'' of the model, which
in turn is defined to be the coefficient of the asymptotic exponential
decay of the two point correlation function of the order parameter.
In $SU(N)\times SU(N)$ chiral models one defines
\begin{equation}
G(x) \;=\; {1\over N} \langle\,{\rm Tr} \,[ U^\dagger (x) U(0) ]\,
\rangle
\label{u1}
\end{equation}
and the true mass gap is
\begin{equation}
\mu \;=\; - \, \lim_{|x|\rightarrow \infty} {\ln G(x)\over |x|}\;\;.
\label{u2}
\end{equation}
It is also possible  to introduce the lattice momentum transform
\begin{equation}
\widetilde{G}(p)\;=\; \sum_x G(x) \exp\left( i{2\pi p\cdot x
\over L}\right)
\label{u3}
\end{equation}
and extract the mass gap from the (imaginary momentum) pole
singularity of $\widetilde{G}(p)$, i.e. by solving the equation
\begin{equation}
\widetilde{G}^{-1}(p=i\mu)\;=\;0\;\;.
\label{u4}
\end{equation}
There are a few well-known (but not necessarily well understood)
results concerning the evaluation of the mass gap.
In particular it is often stated that in the strong coupling regime $G(x)$
``does not exponentiate'' and it is therefore necessary to define the
wall-wall correlation
\begin{equation}
G_w(x_\parallel)\;=\;\sum_{x_\perp} G(x_\perp, x_\parallel)
\label{u5}
\end{equation}
enjoying the exponentiation property, and assume
\begin{equation}
\mu \;=\; - \, \lim_{|x_\parallel |\rightarrow \infty} {\ln G_w(x_\parallel)
\over |x_\parallel |}\;\;.
\label{u6}
\end{equation}
Appendix~\ref{BA} is devoted to a short discussion of this question, which
finds an easy illustration in the context of the exactly solvable Gaussian
model.

In practice, when extracting physical quantities
from Monte Carlo simulations, one typically faces a situation where it is
non-realistic
to solve numerically Eq.~(\ref{u4}), which requires an analytic
continuation to negative $p^2$ of the (usually poorly known) function
$\widetilde{G}^{-1}(p)$. We shall be able to perform this exercise in our
specific study, but this is far from being an easily generalizable
practice.
On the other side, working on a finite lattice often prevents from
exploring a sizeable region where exponentiation may hold with small
errors for arbitrary $\beta$. We therefore found convenient to analyze
explicitly the strong coupling expansion of the Green function $G(x)$ for
finite fixed lattice distance, in order to establish a benchmark both for
exponentiation and for the analysis of the numerical results.

Without entering many details of the (sometimes cumbersome) strong
coupling calculations, we want to sketch the main ingredients and logical
steps before presenting our results. In passing we notice that we might
consider a slightly more general Green's function than Eq.~(\ref{u1}):
\begin{equation}
G_{(r)}\;=\;{\langle\,\chi_{(r)}[ U^\dagger (x) U(0) ]\,\rangle \over
d_{(r)}}\;\;,
\label{u17}
\end{equation}
where $r$ can be restricted, for principal chiral models, to the completely
antisymmetric representations $(1^n;0)$.
The strong coupling expansion is naturally ordered in the length of the
path that we choose in order to connect the point $x=(x_1,x_2)$
with the origin.
In turn, since every path can be decomposed into a sum of shorter paths, we
may in general establish recursive relationships connecting the (fixed
length) contributions to $G(x_1,x_2)$ with the (shorter length)
contributions to Green's functions of nearby points.

The strong coupling character expansion of Green's functions involves, to
lowest orders, only a summation over properly weighted self-avoiding walks.
The effect of ``potentials'' is a higher order contribution that must be
included by considering bifurcating paths ``dressed'' with properly chosen
representations of the link operators.
For all $U(N)$ groups ($N\ge 2$) and $SU(N)$ groups ($N\ge 4$)
the $O\left( \beta^5 \right)$ strong coupling character expansion may be
represented by
\begin{eqnarray}
G(x_1,x_2)\;=&& z_1^{x_1+x_2}\,\Biggl[ C_0(x_1,x_2)\,+\,
C_2(x_1,x_2)z_1^2\,+\,C_4(x_1,x_2)z_1^4\nonumber \\
&&\;\;\;\;\;\;\; + \,A(x_1,x_2)z_1^2\,W_{11}\,+\,
B(x_1,x_2)\,W_{21}\,+\,O\left( z_1^6\right)\Biggr]\;\;,
\label{u28}
\end{eqnarray}
where the quantities $C_{2k}(x_1,x_2)$ represent the number of self-avoiding
walks connecting the origin with the lattice point $x=(x_1,x_2)$ and whose
length is $l=x_1+x_2+2k$, while
the functions $A$ and $B$ satisfy the relationship
\begin{equation}
A(x_1,x_2)\,+\,2B(x_1,x_2)\;=\;
2(x_1+x_2)\,C_0(x_1,x_2)\;\;.
\label{u29}
\end{equation}

In Appendix~\ref{BB} we derived the following results
\begin{equation}
C_0(x_1,x_2)\;=\;
\left( \begin{array}{c} x_1+x_2\\ x_1\end{array}\right)
\;\equiv\;\left( \begin{array}{c} x_1+x_2\\ x_2\end{array}\right)\;\;,
\label{u21}
\end{equation}
\begin{equation}
C_2(x_1,x_2)\;=\;
\left( \begin{array}{c} x_1+x_2\\ x_1\end{array}\right)\,
\left[ {x_1(x_1+1)\over x_2+1}\,+\,{x_2(x_2+1)\over x_1+1}\right]\;\;,
\label{u24}
\end{equation}
\begin{eqnarray}
C_4(x_1,x_2)\;=&&
\left( \begin{array}{c} x_1+x_2\\ x_1\end{array}\right)\;
\Biggl[ x_1x_2 + 2x_1+2x_2+
{x_2(x_2+3)\over x_1+1}\,+\,{x_1(x_1+3)\over x_2+1}\nonumber \\
&&\;\;\;+{(x_2-1)x_2(x_2+1)(x_2+2)\over 2(x_1+1)(x_1+2)}+
{(x_1-1)x_1(x_1+1)(x_1+2)\over 2(x_2+1)(x_2+2)}-
{2x_1x_2\over x_1+x_2}\Biggr]\;\;,
\label{u27}
\end{eqnarray}
\begin{equation}
A(x_1,x_2)\;=\;2{x_1^2+x_2^2\over x_1+x_2}\,C_0(x_1,x_2)\;\;,
\label{u35}
\end{equation}
\begin{equation}
B(x_1,x_2)\;=\; {2x_1x_2\over x_1+x_2}\,C_0(x_1,x_2)\;\;.
\label{u34}
\end{equation}

Writing down and solving recursion equations for the coefficients of the
strong-coupling series for Green's functions is a very powerful but by no
means simple or efficient approach to the evaluation of the mass gap.
In practice we may observe that, to any given order of the strong coupling
expansion, the recursive relations between coefficients imply that only a
finite number of short distance wall-wall correlations for $L\le \bar{L}$
may violate exact exponentiation because of boundary condition effects.
Therefore one may compute the quantities
\begin{equation}
\ln G_w(L+1)\,-\,\ln G_w(L)
\label{u382}
\end{equation}
for the first few values of $L$ until they become constant.
The constants obtained are directly related to the masses by the
relationships
\begin{equation}
\mu_{side}\;=\; \ln G_{side}(L)\,-\,\ln G_{side}(L+1)
\;\;\;\;\;\;\;\;\;\;\;L>{\bar{L}}_s\;\;,
\label{u3825}
\end{equation}
\begin{equation}
{\mu_{diag}\over \sqrt{2}}\;=\; \ln G_{diag}(L)\,-\,\ln G_{diag}(L+1)
\;\;\;\;\;\;\;\;\;\;\;L>{\bar{L}}_d\;\;,
\label{u383}
\end{equation}
where by definition
\begin{equation}
G_{side}(L)\;\equiv\;\sum_{x_2=-\infty}^\infty G(L,x_2)\;\;,
\label{u36}
\end{equation}
\begin{equation}
G_{diag}(L)\;\equiv\;\sum_{x_2=-\infty}^\infty G(L-x_2,x_2)\;\;.
\label{u37}
\end{equation}
It is possible to prove that the alternative definition of $\mu$ based on
solving the equation
\begin{equation}
\widetilde{G}^{-1}(p=i\mu)\;=\;0
\label{u384}
\end{equation}
perturbatively in the strong coupling expansion parameter gives stable
results only after all correlations up to $L={\bar{L}}$
have been included in the Fourier transform of the propagator, and the
result obviously coincides with Eqs.~(\ref{u3825}) and (\ref{u383}).
It is therefore convenient to construct explicitly the Fourier transform
of the inverse propagator and extract the relevant physical parameters by
a direct analysis of this last quantity, making use of the above
considerations in order to establish the accuracy of the computations,
which in general will not coincide with the precision reached in the
evaluation of $\widetilde{G}^{-1}(p)$.

The evaluation of $\widetilde{G}^{-1}(p)$ is dramatically simplified by the
observation that any strong-coupling expanded two-point Green's function
can be unambiguously separated into the form:
\begin{equation}
G(x)\;=\;G_0(x)\;+\;\Delta G(x)\;\;,
\label{m1}
\end{equation}
where $G_0(x)$ is originated by the Fourier transform of the generalized
Gaussian propagator
\begin{equation}
\widetilde{G}_0(p)\;=\;{1\over A_0(z_1)\,-\,2z_1B_0(z_1)\sum_\mu \cos p_\mu
}\;\;,
\label{m2}
\end{equation}
and the functions $A_0(z_1)$ and $B_0(z_1)$ are uniquely determined by the
conditions
\begin{eqnarray}
G_0(0)&=&1\;\;,\nonumber \\
G_0(1,0)&=&1\,-\,E(z_1)\;=\; {1\over 4}{d{\cal F}\over d\beta}\;\equiv
\;\varepsilon (z_1)\;\;,
\label{m3}
\end{eqnarray}
which can be cast into the form:
\begin{eqnarray}
1&=&{1\over A_0} \sum_{n=0}^\infty
\left( \begin{array}{c} 2n\\ n\end{array}\right)^2
\left( {zB_0\over A_0}\right) ^{2n}\;\;,\nonumber \\
zB_0&=&{A_0\,-\,1\over 4\varepsilon }\;\;.
\label{m4}
\end{eqnarray}
{}From the results of the previous section we may extract:
\begin{eqnarray}
\varepsilon (z_1)\;=
&&z_1+2z_1^3+6z_1^5+2z_1^3W_{11}+
+28z_1^7+12z_1^5W_{11}+2z_1W_{31}\nonumber \\
&&+140z_1^9+76z_1^7W_{11}+6z_1^5W_{20}+8z_1^3W_{31}
+2V_{131}+z_1^6V_{100}+O\left(z_1^{11}\right)\;.
\label{m5}
\end{eqnarray}
As a consequence we obtain
\begin{eqnarray}
A_0(z_1)\;=&&
1+4z_1^2+12z_1^4+60z_1^6+16z_1^4W_{11}+
+316z_1^8+96z_1^6W_{11}+16z_1^2W_{31}+1844z_1^{10}\nonumber \\
&&+848z_1^8W_{11}+64z_1^6W_{11}^2
+48z_1^6W_{20}+64z_1^4W_{31}
+16z_1V_{131}+8z_1^7V_{100}+O\left(z_1^{12}\right)\,,
\label{m6}
\end{eqnarray}
\begin{eqnarray}
B_0(z_1)\;=&&
1+z_1^2+7z_1^4+2z_1^2W_{11}+31z_1^6+6z_1^4W_{11}+
+2z_1^2W_{31}+189z_1^{8}+86z_1^6W_{11}\nonumber \\
&&+6z_1^4W_{11}^2
+6z_1^4W_{20}+2z_1^2W_{31}
+2z_1^{-1}V_{131}+z_1^5V_{100}+O\left(z_1^{10}\right)\;.
\label{m7}
\end{eqnarray}
A direct evaluation of all coordinate-space Green's functions
that are non-trivial to $O\left( z_1^{10}\right)$ allows
us to determine $\Delta G(x)$ with the same precision,
since $G_0(x)$ is easily obtained by antitransforming Eq.~(\ref{m2}).
By noticing that the first non-trivial contributions to $\Delta G(x)$ are
$O\left( z_1^6\right)$, it is now relatively easy to evaluate directly
$\widetilde{G}^{-1}(p)$ to $O\left( z_1^{10}\right)$.
We obtained:
\begin{eqnarray}
\widetilde{G}^{-1}(p)\;=&&A(z_1)\,+\,z_1B(z_1)\hat{p}^2
\,+\,z_1^6C(z_1)\hat{p}^2_1\hat{p}^2_2\nonumber \\
&&\,+\,
z_1^8D(z_1)\left( \hat{p}^4_1+\hat{p}^4_2\right)\,+\,
z_1^9E(z_1)\hat{p}^2\hat{p}^2_1\hat{p}^2_2
\,+\,O\left( z_1^{11}\right) \;,
\label{m8}
\end{eqnarray}
where $\hat{p}^2_\mu=4\sin^2
\left( {p_\mu/ 2}\right)$, $\hat{p}^2=\sum_\mu\hat{p}^2_\mu$, and
\begin{eqnarray}
A(z_1)\;=&&A_0(z_1)-4z_1B_0(z_1)-8\Delta^{(1)}z_1^6
\left( 1-4z_1+3z_1^2+6z_1^3-3z_1^4 \right)\nonumber \\
&&-16\Delta^{(2)}z_1^8\left( 1-4z_1+2z_1^2\right)
-8\Delta^{(3)}z_1^9-8\Delta^{(4)}z_1^{10} - 16\Delta^{(5)}z_1^{10}
+O\left( z_1^{11}\right)\;,
\label{m9}
\end{eqnarray}
\begin{eqnarray}
B(z_1)\;=&&B_0(z_1)+4\Delta^{(1)}z_1^5\left( 1-2z_1+2z_1^2+3z_1^3\right)
+8\Delta^{(2)}z_1^7\left( 1-2z_1\right)\nonumber \\
&&+10\Delta^{(3)}z_1^8+8\Delta^{(4)}z_1^9+8\Delta^{(5)}z_1^9
+O\left( z_1^{10}\right)\;,
\label{m10}
\end{eqnarray}
\begin{equation}
C(z_1)\;=\;-2\Delta^{(1)}-4\Delta^{(2)}z_1^2
-8\Delta^{(3)}z_1^3-4\Delta^{(5)}z_1^4
+O\left( z_1^{5}\right)\;,
\label{m11}
\end{equation}
\begin{equation}
D(z_1)\;=\;-2\Delta^{(1)}-2\Delta^{(3)}z_1-2\Delta^{(4)}z_1^2
+O\left( z_1^{3}\right)\;,
\label{m12}
\end{equation}
\begin{equation}
E(z_1)\;=\;\Delta^{(3)}+O\left( z_1^2\right)\;.
\label{m13}
\end{equation}
We have defined the following combinations of potentials:
\begin{equation}
\Delta^{(1)}\;=\;z_1^{-4}W_{21}-2z_1^{-2}W_{11}-1\;\;,
\label{m14}
\end{equation}
\begin{equation}
\Delta^{(2)}\;=\;-z_1^{-6}W_{31}+4z_1^{-4}W_{21}
+z_1^{-2}W_{10}-7z_1^{-2}W_{11}-4
\label{m15}
\end{equation}
\begin{equation}
\Delta^{(3)}\;=\;z_1^{-2}\widetilde{W}_{11}
+2z_1^{-6}W_{31}-4z_1^{-4}W_{21}
-2z_1^{-2}W_{10}+4z_1^{-2}W_{11}+1
\label{m16}
\end{equation}
\begin{eqnarray}
\Delta^{(4)}\;=&&
2z_1^{-6}\widetilde{W}_{21}+z_1^{-6}W_{111}-z_1^{-4}\widetilde{W}_{11}
-4z_1^{-6}W_{31}+z_1^{-8}W_{41}+16z_1^{-4}W_{21}\nonumber \\
&&-2z_1^{-4}W_{20}-3z_1^{-4}W_{11}^2-26z_1^{-2}W_{11}+4z_1^{-2}W_{10}-9\;\;,
\label{m17}
\end{eqnarray}
\begin{eqnarray}
\Delta^{(5)}\;=&&
z_1^{-5}V_{110}+{1\over 2}z_1^{-10}V_{221}-
z_1^{-9}V_{131}-z_1^{-3}V_{100}-z_1^{-4}\widetilde{W}_{11}
+z_1^{-8}W_{41}-8z_1^{-6}W_{31}\nonumber \\
&&+{55\over 2}z_1^{-4}W_{21}-3z_1^{-4}W_{20}-
{9\over 2}z_1^{-4}W_{11}^2-53z_1^{-2}W_{11}+
13z_1^{-2}W_{10}-{53\over 2}\;\;.
\label{m18}
\end{eqnarray}
Eq.~(\ref{m8}) is a rather compact collection of physically
relevant results concerning the strong coupling regime of the models
under investigations.
We notice that the nearest-neighbor Gaussian structure of the propagator
starts being violated to $O\left( z_1^6\right)$ in diagonal correlations
and only to $O\left( z_1^8\right)$ in side correlations.
We therefore
expect a substantial agreement of the ratio $\xi_G/\xi_w$,
where $\xi_G^2\equiv\langle \,x^2\,\rangle$ and $\xi_w\equiv 1/\mu_{side}$,
with its Gaussian
value even for not too strong values of the coupling
(as we will see in Section~\ref{numericalresults}).

By standard arguments we can establish relationships between the
propagator, the susceptibility and its second moment:
\begin{eqnarray}
\chi&=&\sum_{x_1,x_2}G(x_1,x_2)\;=\;
G_{side}(0)\,+\,2\sum_{L=1}^\infty G_{side}(L)\nonumber \\
&=&
G_{diag}(0)\,+\,2\sum_{L=1}^\infty G_{diag}(L)\;=\;{1\over A(z_1)}\;\;,
\label{m19}
\end{eqnarray}
\begin{eqnarray}
\chi\;\langle \,x^2\,\rangle &=&
{1\over 4}\sum_{x_1,x_2}(x_1^2+x_2^2)G(x_1,x_2)\;=\;
\sum_{L=1}^\infty L^2 G_{side}(L)\nonumber \\ &=&
{1\over 2}\sum_{L=1}^\infty L^2 G_{diag}(L)\;=\;
{z_1B(z_1)\over A(z_1)^2}\;\;.
\label{m20}
\end{eqnarray}
Hence we obtain
\begin{equation}
M^2_{G}\;\equiv\;{1\over \langle \,x^2\,\rangle } \;=\;
{A(z_1)\over z_1B(z_1)}\;\;,
\label{m21}
\end{equation}
\begin{equation}
Z_G\;=\;{1\over z_1 B(z_1)}\;\;.
\label{m22}
\end{equation}
Moreover by solving appropriate algebraic equations we find
\begin{equation}
M^2_{{side}}\;=\;2 \,\left( {\rm ch} \,\mu_{side}\;-\,1\right)\;\simeq\;
M^2_{G}\,+\,{z_1^5A(z_1)^2 D(z_1)\over B(z_1)^3}\,+\,...
\label{m23}
\end{equation}
\begin{equation}
M^2_{{diag}}\;=\;4 \,\left( {\rm ch} \,{\mu_{diag}\over
\sqrt{2}}\;-\;1\right)
\;\simeq\;M^2_{G}\,+\,{A(z_1)^2 \over B(z_1)^3}\left( {z_1^5D(z_1)\over
2}+{z_1^3C(z_1)\over 4}\right) - {z_1^5A(z_1)^3E(z_1)\over 4B(z_1)^4}\,+\,...
\label{m24}
\end{equation}
By substituting our explicit results we find
\begin{eqnarray}
&&\chi\;=\; 1+4z_1+12z_1^2+36z_1^3+100z_1^4+
\Bigl(284+8z_1^{-2}W_{11}\Bigr)z_1^5+
\Bigl(788+48z_1^{-2}W_{11}+8\Delta^{(1)}\Bigr)z_1^6 \nonumber \\
&&+\Bigl(2204+216z_1^{-2}W_{11}+8z_1^{-6}W_{31}+32\Delta^{(1)}\Bigr)z_1^7+
\Bigl(6068+800z_1^{-2}W_{11}+48z_1^{-6}W_{31}+88\Delta^{(1)}\nonumber \\
&&\;+16\Delta^{(2)}\Bigr)z_1^8
+\Bigl(16820+8z_1^{-9}V_{131}+4z_1^{-4}V_{100}+2904z_1^{-2}W_{11}+
24z_1^{-4}W_{11}^2 +24z_1^{-4}W_{20}\nonumber \\
&&\;+200z_1^{-6}W_{31}
+304\Delta^{(1)}+64\Delta^{(2)}+8\Delta^{(3)}\Bigr)z_1^9
+\Bigr(46172+48z_1^{-9}V_{131}+24z_1^{-4}V_{100}+9840z_1^{-2}W_{11}
\nonumber \\
&&\;+192z_1^{-4}W_{11}^2+144z_1^{-4}W_{20}+704z_1^{-6}W_{31}
+1000\Delta^{(1)}
+160\Delta^{(2)}+64\Delta^{(3)}+8\Delta^{(4)}+16\Delta^{(5)}\Bigr)
z_1^{10}\nonumber \\
&&+O\left(z_1^{11}\right)\;,
\label{m25}
\end{eqnarray}
\begin{eqnarray}
&&z_1M^2_{G}\;=\;
1-4z_1+3z_1^2+\Bigl(2-2z_1^{-2}W_{11}\Bigr)z_1^4-4\Delta^{(1)}z_1^5+
\Bigl(6+6z_1^{-2}W_{11}-2z_1^{-6}W_{31}+16\Delta^{(1)}\Bigr)z_1^6 \nonumber \\
&&-\Bigl(16\Delta^{(1)}+8\Delta^{(2)}\Bigr)z_1^7
+\Bigl(14-2z_1^{-9}V_{131}-z_1^{-4}V_{100}-4z_1^{-2}W_{11}-
2z_1^{-4}W_{11}^2-6z_1^{-4}W_{20}
\nonumber \\
&&\;+10z_1^{-6}W_{31}+4\Delta^{(1)}+32\Delta^{(2)}-10\Delta^{(3)}\Bigr)z_1^8
+\Bigl(12\Delta^{(1)}-16\Delta^{(2)}+32\Delta^{(3)}-8\Delta^{(4)}\nonumber \\
&&\;-8\Delta^{(5)}+16z_1^{-2}W_{11}\Delta^{(1)}\Bigr)z_1^9
+O\left( z_1^{10}\right)\;,
\label{m26}
\end{eqnarray}
\begin{eqnarray}
&&z_1Z_G\;=\;
1-z_1^2+\Bigl(-6-2z_1^{-2}W_{11}\Bigr)z_1^4-4\Delta^{(1)}z_1^5+
\Bigl(-18-2z_1^{-2}W_{11}-2z_1^{-6}W_{31}+8\Delta^{(1)}\Bigr)z_1^6
\nonumber \\
&&-8\Delta^{(2)}z_1^7
+\Bigl(-98-2z_1^{-9}V_{131}-z_1^{-4}V_{100}-52z_1^{-2}W_{11}-2z_1^{-4}W_{11}^2
-6z_1^{-4}W_{20}+2z_1^{-6}W_{31}
\nonumber \\
&&\;-28\Delta^{(1)}+16\Delta^{(2)}-10\Delta^{(3)}\Bigr)z_1^8
 +\Bigl(60\Delta^{(1)}+16\Delta^{(2)}-8\Delta^{(4)}
-8\Delta^{(5)}+16z_1^{-2}W_{11}\Delta^{(1)}\Bigr)z_1^9
\nonumber \\
&&+O\left( z_1^{10}\right)\;,
\label{m27}
\end{eqnarray}
\begin{eqnarray}
&&\mu_{side}\;=\;-\ln z_1-2z_1-\Bigl( 2+2z_1^{-2}W_{11}\Bigr)z_1^4
-\Bigl( {22\over 5}+4z_1^{-2}W_{11}+4\Delta^{(1)}\Bigr)z_1^5\nonumber \\
&&-\Bigl( 4 + 2z_1^{-6}W_{31}-6\Delta^{(1)}\Bigr) z_1^6-
\Bigl( {86\over 7}+4z_1^{-2}W_{11}+4z_1^{-6}W_{31}-16\Delta^{(1)}
+8\Delta^{(2)}+2\Delta^{(3)}\Bigr) z_1^7\nonumber \\
&&-\Bigl( 28 + 32z_1^{-2}W_{11}+4z_1^{-4}W_{11}^2+6z_1^{-4}W_{20}
-4z_1^{-6}W_{31}+2z_1^{-9}V_{131}+z_1^{-4}V_{100}+
+28\Delta^{(1)}
\nonumber \\
&&\;-16\Delta^{(2)}-2\Delta^{(3)}+2\Delta^{(4)}\Bigr) z_1^8
-2\Delta^{(6)}z_1^8
+O\left( z_1^{9}\right)\;,
\label{m28}
\end{eqnarray}
\begin{eqnarray}
{\mu_{diag}\over \sqrt{2}}\;=&& -\ln 2z_1-z_1^2
-\Bigl( {5\over 2} + 2z_1^{-2}W_{11}+{1\over 2}\Delta^{(1)}\Bigr) z_1^4
\nonumber \\
&&-\Bigl( {25\over 3} + 4z_1^{-2}W_{11}+2z_1^{-6}W_{31}
-2\Delta^{(1)}+\Delta^{(2)}+{1\over 4}\Delta^{(3)}\Bigr) z_1^6
+O\left( z_1^{7}\right)\;.
\label{m29}
\end{eqnarray}
The contribution proportional to
\begin{equation}
\Delta^{(6)}\;\equiv\;z_1^{-6}W_{31}-2z_1^4W_{21}+z_1^2W_{11}
\label{u395}
\end{equation}
in Eq.~(\ref{m28}) was obtained by evaluating the $O\left( z_1^{11}\right)$
contribution to $G(3,0)$ and could not have been predicted from the
knowledge of the $O\left( z_1^{10}\right)$ contributions to
$\widetilde{G}^{-1}(p)$: this is in accord with our previously discussed
considerations.
Further useful results are:
\begin{eqnarray}
&&\ln G_{side}(L)\;=\; L\ln z_1+ 2(L+1)z_1+{2\over 3}(L+7)z_1^3
+2(L+2)z_1^4+{2\over 5}(L+51)z_1^5+2Lz_1^2W_{11}\nonumber \\
&& -4(L-1)z_1^3W_{11}+
4Lz_1W_{21}+2(5L+9)z_1^6+4(3L+1)z_1^4W_{11}
-2(3L+1)z_1^2W_{21}+2LW_{31}
\nonumber \\
&&
+2z_1^6\Delta^{(1)}\delta_{L,0}
+ {4\over 7}(172-3L)z_1^7+
4(L+1)z_1^5W_{10}-12(L-1)z_1^5W_{11}+4(2L+3)z_1^3W_{21}
\nonumber \\
&&+2(L-1)z_1^3\widetilde{W}_{11}
+2z_1^7\Bigl(\Delta^{(3)}-2\Delta^{(1)}\Bigr)
\delta_{L,0}+22(2L+5)z_1^8-4(L+4)z_1^6W_{10}+2(15L+28)z_1^6W_{11}
\nonumber \\
&&-2(L-3)z_1^4W_{11}^2 +2(L+2)z_1^4W_{20} - 20z_1^4W_{21}
+2(L+6)z_1^2W_{31}+2(L-1)W_{41}-2(2L-3)z_1^4\widetilde{W}_{11}
\nonumber \\
&&+4(L-1)z_1^2\widetilde{W}_{21}
 +2(L-1)z_1^2W_{111}+2z_1^{-1}LV_{31}+z_1^5LV_{100}+
2z_1^8\Delta^{(6)} \delta_{L,1}\nonumber \\
&&+2z_1^8\Bigl( 2\Delta^{(6)}
+\Delta^{(4)}-2\Delta^{(3)}\Bigr) \delta_{L,0}
+O\left( z_1^9\right)\;,
\label{u39}
\end{eqnarray}
\begin{eqnarray}
&&\ln G_{diag}(L)\;=\; L\ln 2z_1+ (L+4)z_1^2+
(2L+{25\over 2})z_1^4 + (L+1)z_1^2\,W_{11}
+{1\over 2}(L-1)\,W_{21}
\nonumber \\
&&+{1\over 2}z_1^4\Delta^{(1)}
\delta_{L,0}+{754+79L\over 12}z_1^6+ {L\over 2}z_1^4W_{10}
+(2L+9)z_1^4W_{11}+ (L+4)z_1^2W_{21}+{3\over 2}LW_{31}
\nonumber \\
&&+{L-2\over 4}z_1^2\widetilde{W}_{11}
+{1\over 4}z_1^6\Delta^{(3)}
\delta_{L,1}
+ {1\over 2}z_1^6\Bigl(\Delta^{(3)}+2\Delta^{(2)}-4\Delta^{(1)}\Bigr)
\delta_{L,0}
+O\left( z_1^8\right)\;.
\label{u40}
\end{eqnarray}

The results we have obtained are expressed in the language of the character
expansion. However within the precision of our expansion it is possible,
for all $U(N)$ and $SU(N)$ groups with $N$ sufficiently large, to convert the
results
into a standard strong coupling series by the following replacements
(dictated by our previous analysis)
\begin{equation}
z_1\;\simeq\;\beta\;\;,
\label{u61}
\end{equation}
\begin{equation}
z_1^{-2}W_{11}\;\simeq\;0\;\;,
\label{u62}
\end{equation}
\begin{equation}
z_1^{-2}W_{10}\;\simeq\;2z_1^{-4}W_{21}\;\simeq\;
{4N^2\over N^2-1}\;\goto_{N\rightarrow \infty}4\;\;,
\label{u63}
\end{equation}
\begin{equation}
z_1^{-4}W_{20}\;\simeq\;2z_1^{-6}W_{31}
\;\simeq\;{2N^2(5N^2-2)\over (N^2-1)^2} \;\goto_{N\rightarrow \infty}10\;\;,
\label{u632}
\end{equation}
\begin{equation}
z_1^{-8}W_{41}
\;\simeq\;{N^2(9N^4-6N^2+2)\over (N^2-1)^3} \;\goto_{N\rightarrow \infty}9\;\;,
\label{u633}
\end{equation}
\begin{equation}
z_1^{-2}\widetilde{W}_{11}\;\simeq\; {N^2(7N^2-2)\over (N^2-1)^2}
\;\goto_{N\rightarrow \infty}7\;\;,
\label{u651}
\end{equation}
\begin{equation}
z_1^{-6}\widetilde{W}_{21}\;\simeq\;
{N^2(11N^4-6N^2+2)\over (N^2-1)^3}
\;\goto_{N\rightarrow \infty} 11\;\;,
\label{u652}
\end{equation}
\begin{equation}
z_1^{-6}W_{111}\;\simeq\; {-4N^4\over (N^2-1)^2}
\;\goto_{N\rightarrow \infty}-4\;\;,
\label{u653}
\end{equation}
\begin{equation}
z_1^{-4}V_{100}\;\simeq\;0\;\;,
\label{u72}
\end{equation}
\begin{equation}
z_1^{-5}V_{110}\;\simeq\;0\;\;,
\label{u69}
\end{equation}
\begin{equation}
z_1^{-9}V_{131}\;\simeq\;0\;\;,
\label{u71}
\end{equation}
\begin{equation}
z_1^{-10}V_{221}\;\simeq\; {8N^4(3N^2-2)\over (N^2-1)^2(N^2-4)}
\;\goto_{N\rightarrow \infty}24\;\;.
\label{u70}
\end{equation}
In Appendix~\ref{BC} we give explicitly the $N=\infty$ strong coupling
series of some relevant quantities considered in this section.

\section{Weak coupling expansion}
\label{weakcoupling}

For both continuum and lattice, at low temperature, the perturbative
expansion is performed by setting
\begin{equation}
U\;=\;e^{iA}\;\;, \;\;\;\;\;A=\sum_a T_a A_a\;\;,
\label{chvar}
\end{equation}
($T_a$ are the generators of the $SU(N)$ group and $A_a$ are
$N^2-1$ real fields) and expanding $U$ in powers of $A$.
The above change of variables introduces an additional
ill-defined determinant in the partition function, indeed \cite{Kawai}
\begin{eqnarray}
[dU]&=&K\, \prod_a [dA_a] e^{-S_m}\;\;, \nonumber \\
S_m &=& -{1\over 2} \sum_n {\rm Tr}\;\ln {2(1-\cos \widetilde{A}_n)\over
\widetilde{A}_n^2}\;\;,
\label{measure}
\end{eqnarray}
where $\widetilde{A}_{bc}=\sum_a if_{bac} A_a$ and $K$ is an irrelevant
constant.
In dimensional regularization, the measure term does not contribute,
as a consequence of the rule $\int d^dk=0 \Leftrightarrow \delta^d(0)=0$,
where $d$ is the space dimension.

Short weak coupling series for the free-energy density of $U(N)$ and $SU(N)$
chiral
models on the lattice were presented in Ref. \cite{Brihaye-Rossi}.
We calculated the energy density up to three loops finding
\begin{equation}
E\;=\; 1\;-\;\langle \;{1\over N}
{\rm Re} \;{\rm Tr} \,[ U_n U^\dagger_{n+\mu} ]\;
\rangle \;=\; {N^2-1\over 8N^2\beta}\left[ 1\,+\,{a_1\over\beta}\,+\,
{a_2\over \beta^2}\,+\,...\;\right]
\label{WCenergy}
\end{equation}
where
\begin{eqnarray}
a_1&=& {N^2-2\over 32N^2} \;\;\;,\nonumber \\
a_2&=& {3N^4-14N^2+20\over 768N^4}\,+\, {N^4-4N^2+12\over 64N^4}Q_1
\,+\, {N^4-8N^2+24\over 64N^4}Q_2\;\;\;,
\end{eqnarray}
$Q_1$ and $Q_2$ being numerical constants:
\begin{eqnarray}
Q_1&=&\int {d^2k\over (2\pi)^2}
\; \left[ \int {d^2p\over (2\pi)^2}
\left( {\widehat{k}^2 \over \widehat{p}^2 \widehat{p+k}^2}
-{2\over \widehat{p}^2} \right)\right]^2\;\;,\nonumber \\
Q_2&=& \sum_{\mu,\nu} \left[ \int \delta (\sum_1^4 k_i)
{ \widehat{(k_1+k_2)}^2_\mu \widehat{(k_1+k_3)}^2_\nu\over \widehat{k}^2_1\,
\widehat{k}^2_2 \,\widehat{k}^2_3 \,\widehat{k}^2_4} \right]
\;-\;2\left[ \int {1\over
\widehat{k}^2}\right]^2\;\;.
\end{eqnarray}
$Q_1=0.0958876$ and $Q_2=-0.0670$.

We calculated
\begin{equation}
G(T,x)\;=\; {\int [dU] \;{1\over N} {\rm Re}\,{\rm Tr}\,[U(0) U(x)^\dagger]\;
\exp (-S/T)\over \int [dU] \;\exp (-S/T)}
\label{greenfuctp}
\end{equation}
in perturbation theory, and in two regularization schemes:
dimensional and lattice regularizations.
In dimensional regularization and in the $x$-space
\begin{equation}
G_D(T,x,\epsilon)\;=\; 1\;+\;{N^2-1\over 2N} \,{T\over (2\pi)^d S_d}\,
{1\over \epsilon x^\epsilon}\;+\;
{N^2-1\over 2N}\,{N^2-2\over 8N}\,{T^2\over (2\pi)^{2d} S_d^2}\,
{1\over \epsilon^2 x^{2\epsilon}}\;+\,...
\label{dimenx}
\end{equation}
where $S_d=\left[ (2\pi)^{d/2} 2^{\epsilon/2} \Gamma(1+\epsilon/2)\right]^{-1}$
and $\epsilon =d-2$.
In the $p$-space
\begin{equation}
\widetilde{G}_D(T,p,\epsilon)\;=\;{N^2-1\over 2N} {T\over p^2}\,\left[
1\,+\, {N^2-2\over 4N}T S_d {p^\epsilon\over \epsilon}\;+
T^2 S_d^2 \,p^{2\epsilon}\left( {-N^2+2\over 16N^2}{1\over \epsilon^2}
-{N^2\over 64}{1\over \epsilon} + {3N^2\over 128} \right)\,+\,...\right]
\label{dimenp}
\end{equation}

On the lattice we obtained (neglecting $O\left( a^2 \right)$ terms)
\begin{equation}
G_L(T,x,a)\;=\; 1\;+\;{N^2-1\over 2N} \,T \,E(x/a)\;+\;
{N^2-1\over 2N}\,{N^2-2\over 16N}\,T^2 E(x/a)\Bigl(1+2E(x/a)\Bigr)\;+...\,\;
\label{lattx}
\end{equation}
where
\begin{equation}
E(x/a)\;=\; {1\over 2\pi} \left( \ln a/x - \gamma_E - {3\over 2}\ln
2\right)\;\;;
\end{equation}
\begin{eqnarray}
&&\widetilde{G}_L(T,p,a)\;=\;{N^2-1\over 2N} {T\over p^2}\;\Biggl[
1\;+\; {N^2-2\over 4N}T \left( B(pa)+{1\over 4}\right)\;+\nonumber \\
&&T^2 \left( {-N^2+2\over 16N^2}B(pa)^2 +
{N^4-4N^2+4\over 32N^2} B(pa) -{N^2\over 32} C(pa)+
{5N^4-25N^2+30\over 768N^2} +{N^2\over 32}G_1\right)
+...
\label{lattp}
\end{eqnarray}
where
\begin{eqnarray}
B(pa)&=& {1\over 2\pi} \left(\ln pa - {5\over 2}\ln 2\right)\;\;,\nonumber \\
C(pa)&=& {1\over (2\pi)^2} \left(\ln pa - {5\over 2}\ln 2
-{1\over 2}\right)\;\;,
\end{eqnarray}
and $G_1$ is a numerical constant: $G_1= 0.04616363$
\cite{Falcioni,Campo-Rossi}.
The equivalence of the $SU(2)\times SU(2)$ chiral model to the $O(4)$
$\sigma$ model allows a check of these expressions, indeed for $N=2$ they must
give
(and indeed they do) the same results  obtained for the corresponding
correlation functions in  the standard lattice $O(4)$
$\sigma$ model \cite{Falcioni}.

On the lattice, neglecting $O\left( a^2 \ln^n a\right)$ terms,
the correlation function $\widetilde{G}_L(T,p,a)$ satisfies
the renormalization group equation
\begin{equation}
\left[ -a{\partial \over \partial a} \,+\, \beta_L(T){\partial\over \partial T}
\,+\,\gamma_L(T)\right] \widetilde{G}_L(T,p,a) \;=\;0\;\;,
\label{renormeq}
\end{equation}
where the functions $\beta_L(T)$ and $\gamma_L(T)$ tell us how the temperature
and the field $U$ should vary with the lattice spacing $a$ to keep
the renormalized quantities fixed:
\begin{equation}
\beta_L(T) \;\equiv\; -a {d\over da}T\;=\;-\,b_0\,T^2\,-\,b_1\,T^3\,
-\,b_{2_L}\,T^4\,+\,...
\label{beta}
\end{equation}
\begin{equation}
\gamma_L(T) \;\equiv\; a {d\over da}\ln
Z_U\;=\;\gamma_1\,T\,+\,\gamma_{2_L}\,T^2\,+\,\gamma_{3_L}\,T^3\,+\,...
\label{gamma}
\end{equation}
$Z_U$ is the function entering the renormalization of the two point
function
\begin{equation}
G_R(t,x,\mu) \;=\; Z_U(T,a\mu)^{-1}\;G_L(T,x,a)\;\;,
\label{renormG0}
\end{equation}
where $t$ is the renormalized coupling connected to the temperature by the
relation $T=tZ_t$, and $\mu$ is an energy scale.
In Eqs.\ (\ref{beta}) and (\ref{gamma}) $b_0$, $b_1$ and $\gamma_1$ are
universal
coefficients independent of the regularization scheme,
and appearing also in the renormalization group equations giving
the behavior of the renormalized quantities
when varying $\mu$ keeping the bare quantities fixed.
\begin{eqnarray}
b_0&=&{N\over 8\pi}\;\;\;,\;\;\;\; b_1\;=\; {N^2\over
128\pi^2}\;\;\;,\nonumber \\
\gamma_1&=&{N^2-1\over 4N\pi} \;\;\;.
\end{eqnarray}

The coefficients $b_{2_L}$, $\gamma_{2_L}$ and $\gamma_{3_L}$ can be calculated
using the procedure described in Ref.\ \cite{Falcioni}.
We determine the renormalized functions $Z^{\overline {MS}}_t(T,a\mu)$
and $Z^{\overline {MS}}_U(T,a\mu)$ that satisfy the equations
\begin{eqnarray}
&&G_R^{\overline {MS}}(t,x,\mu) \;=\; Z^{\overline
{MS}}_U(T,a\mu)^{-1}\;G_L(T,x,a)\;\;,\nonumber \\
&&T=Z^{\overline {MS}}_t(T,a\mu)\,t\;\;,
\label{renormG_dim}
\end{eqnarray}
where  $t$ and $G_R^{\overline {MS}}(t,x,\mu)$
are respectively the coupling and
the correlation function renormalized in the ${\overline {MS}}$ scheme.
Renormalizing  in the ${\overline {MS}}$ scheme the expressions
(\ref{dimenx}) and (\ref{dimenp}) we find
\begin{eqnarray}
&&G_R^{\overline {MS}}(t,x\mu=2e^{-\gamma_E})\;=\;1\;+\;O(t^3)\;\;,\nonumber
\\ &&\widetilde{G}_R^{\overline {MS}}(t,{p\over \mu}=1)\;=\;
{N^2-1\over 2N}{t\over p^2}\,\left[ 1\,+\,{3N^2\over 128}{t^2\over
(2\pi)^2} \,+\,O(t^3)\right]\;\;,
\label{renormG}
\end{eqnarray}
where $\gamma_E$ is the Eulero constant.
Then by imposing Eq.\ (\ref{renormG_dim}) we obtain
\begin{equation}
Z^{\overline {MS}}_t(T,a\mu)\;=\;
1\;+\;L_1\,T
\;+\;L_2\,T^2\;+\;O(T^3)\;\;,
\label{zetat}
\end{equation}
where
\begin{eqnarray}
L_i&=&c_i\left( \ln a\mu \,+\, d_i\right)\;\;, \nonumber \\
c_1&=&b_0\;\;,\nonumber \\
d_1&=&-\,{5\over 2}\ln 2 \,-\,\pi{N^2-2\over 2N^2}\;\;,\nonumber \\
c_2&=&b_1\;\;,\nonumber \\
d_2&=& -{5\over 2}\ln 2 \,+\,
{1\over 4}\,-\,\pi^2\left( {2N^4-13N^2+18\over 6N^4} +
4G_1\right)\;\;;
\label{coeffzetat}
\end{eqnarray}
\begin{equation}
Z^{\overline {MS}}_U(T,a\mu)\;=\;
1\,+\,M_1 T \,+\, \left[ {1\over 2} \left( 1+{b_0\over \gamma_1}\right)M_1^2
-M_1 L_1\right] T^2 \,+\,O(T^3)\;\;,
\label{zetau}
\end{equation}
where
\begin{eqnarray}
M_1&=& e_1 \left( \ln a\mu + f_1\right)\;\;,\nonumber \\
e_1&=& \gamma_1 \;\;,\nonumber \\
f_1&=&- {5\over 2} \ln 2\;\;.
\label{coeffzetau}
\end{eqnarray}
The ratio of the $\Lambda$-parameters $\Lambda_{\overline {MS}}$ and
$\Lambda_L$ is given by \cite{ShigKogut}
\begin{equation}
{\Lambda_{\overline {MS}}\over \Lambda_L}\;=\; \exp (-d_1)\;=\;
\sqrt{32} \exp \left(
\pi {N^2-2\over 2N^2} \right)\;\;\;.
\label{Lratio1}
\end{equation}
$\gamma_{2_L}$ is easily obtained from  Eq.\ (\ref{gamma}):
\begin{equation}
\gamma_{2_L}\;=\;b_0\gamma_1(f_1-d_1)\;=\; {(N^2-1)(N^2-2)\over 64\pi N^2}\;\;.
\end{equation}

By definition, when keeping the bare quantities $a$ and $T$ fixed
we must have
\begin{equation}
\mu{d\over d\mu} t \;=\;-t\,\mu{d\over d\mu} \ln Z^{\overline
{MS}}_t(T,a\mu)
\;=\; \beta_{\overline {MS}}(t)\;=\;-b_0t^2-b_1t^3-b_{2_{\overline
{MS}}}t^4+...\;,
\label{betams}
\end{equation}
and
\begin{equation}
\mu {d\over d\mu} \ln Z^{\overline {MS}}_U(T,a\mu)\;=\;\gamma_{\overline
{MS}}(t)\;=\; \gamma_1 t+\gamma_{2_{\overline
{MS}}}t^2+\gamma_{3_{\overline {MS}}}t^3+...\;,
\label{gammams}
\end{equation}
where $\beta_{\overline{MS}}(t)$ and $\gamma_{\overline{MS}}(t)$ are
respectively
the $\beta$ function and the anomalous dimension of the field $U$ in the
${\overline{MS}}$ renormalization scheme.
On the other hand, deriving with respect to $a$, keeping the renormalized
quantities $\mu$ and $t$ fixed,
we must obtain the $\beta$ function and the $U$ field anomalous dimension
defined in Eqs.~(\ref{beta}) and (\ref{gamma}):
\begin{equation}
\beta_{L}(T)\;=\;-T\,a{d\over da} \ln Z^{\overline
{MS}}_t(T,a\mu)
\label{betaLL}
\end{equation}
and
\begin{equation}
\gamma_{L}(T)\;=\;a {d\over da} \ln Z^{\overline {MS}}_U(T,a\mu)
\label{gammaLL}
\end{equation}
By comparing the perturbative expansions of Eqs.~(\ref{betams}) and
(\ref{gammams}) with those of Eqs.~(\ref{betaLL}) and (\ref{gammaLL}),
the coefficients $b_{2_L}$ and $\gamma_{3_L}$ can be written in terms
of the quantities introduced in Eq.\ (\ref{zetat}) and the corresponding
coefficients $b_{2_{\overline {MS}}}$, $\gamma_{2_{\overline {MS}}}$
and $\gamma_{3_{\overline {MS}}}$,
which have been already calculated \cite{Hikami}
\begin{equation}
b_{2_{\overline {MS}}}\;=\;{3N^3\over 512}{1\over (2\pi)^3}\;\;,
\label{MScoeff}
\end{equation}
and \cite{Stone}
\begin{eqnarray}
\gamma_{2_{\overline {MS}}}&=&0\;\;,\nonumber \\
\gamma_{3_{\overline {MS}}}&=&{3N(N^2-1)\over 256}{1\over (2\pi)^3}\;\;.
\label{MScoeffg}
\end{eqnarray}
Indeed
\begin{equation}
b_{2_L} \;=\; b_{2_{\overline {MS}}}\,+\,b_0\,b_1\,(d_1\,-\,d_0)\;\;,
\label{b21}
\end{equation}
and
\begin{equation}
\gamma_{3_L} \;=\; \gamma_{3_{\overline {MS}}}\,+\,
\gamma_1\left[ b_0^2\left( d_1-f_1\right)^2 - b_1\left(
d_2-f_1\right)\right]\;\;.
\label{gamma3l}
\end{equation}
In particular we find
\begin{equation}
b_{2_L}\;=\; {1\over (2\pi)^3} \,{N^3\over 128}\left[
1+\pi{N^2-2\over 2N^2}-\pi^2\left( {2N^4-13N^2+18\over 6N^4} +
4G_1\right)\right]\;\;.
\label{b2L}
\end{equation}
Having calculated $b_{2_L}$, we can evaluate the first perturbative
correction to the two loop relationship between the lattice scale
$\Lambda_L$ and the temperature
\begin{eqnarray}
\Lambda_{L} & = & \left( b_0 T \right) ^{-b_1/b_0^2} \exp \left(
-{1\over b_0T} \right)\left[ 1\,+\,{b_1^2-b_0b_{2_L}\over b_0^3}T\,+\,O\left(
T^2\right)\right]\nonumber \\
&=&\left( 8\pi\beta\right)^{1/2}\,e^{-8\pi\beta}
\left[ 1+
{b_1^2-b_0b_{2_L}\over Nb_0^3}\beta^{-1} + O\left( \beta^{-2}\right)
\right]\;\;\;.
\label{ASYSC}
\end{eqnarray}
In the large $N$ limit
\begin{equation}
{b_1^2-b_0b_{2_L}\over Nb_0^3}\goto_{N\rightarrow \infty} 0.0605095\;\;.
\label{Nlimdl}
\end{equation}

In order to solve Eq.\ (\ref{renormeq}), let's introduce
the dimensionless function $H_L(T,pa)=p^2\widetilde{G}_L(T,p,a)$.
A formal solution of Eq.\ (\ref{renormeq}) is given by
\begin{equation}
H_L(T,pa)\;=\; H_L(\Theta,1)\; \exp \left( \int^{\Theta}_T
dz\,{\gamma_L(z)\over
\beta_L(z)} \right)\;\;,
\label{h1}
\end{equation}
where $\Theta\equiv \Theta(T,pa)$ satisfies the equation
\begin{equation}
y{\partial \over \partial y} \Theta (T,y)\;=\;\beta_L(\Theta )\;\;.
\label{Tr}
\end{equation}
Defining the function
\begin{eqnarray}
z(T)&\equiv& \exp \left( \int dT\,{\gamma_L(T)\over
\beta_L(T)} \right)\nonumber \\
&=& T^{-\gamma_1/b_0} \Biggl[ 1 \;+\; {\gamma_1 b_1- \gamma_{2_L} b_0\over
b_0^2}\,T \nonumber \\
&&+\;
{\gamma_{2_L}b_1b_0^2 + \gamma_1 (b_{2_L}b_0-b_1^2)b_0
-\gamma_{3_L} b_0^3   + (\gamma_1b_1-\gamma_{2_L} b_0)^2\over 2\,b_0^4}\,T^2
\;+\;O\left(T^3\right)\Biggr]\,,
\label{zfunction}
\end{eqnarray}
$H_L(T,pa)$ can be rewritten in the form
\begin{equation}
H_L(T,pa)\;=\; z(T)^{-1}\;z(\Theta )\, H_L(\Theta ,1)\;\;.
\label{h2}
\end{equation}
Solving perturbatively Eq.\ (\ref{Tr}) we find
\begin{equation}
\Theta \;=\;{1\over b_0 \,u }\left[ 1\,-\,
{b_1\ln u\over b_0^2\,u} \,+\, {b_1^2 \ln u (\ln u -1) + b_{2_L}b_0-b_1^2\over
b_0^4 \,u^2}\,+\,O\left({\ln^3 u\over u^3}\right)\right]\;\;,
\label{Tr2}
\end{equation}
where $u=\ln (p/\Lambda_L)$.
The perturbative expansion of $H_L(\Theta ,1)$ can be obtained from Eq.\
(\ref{lattp}).

In order to get a more accurate description of the approach
to asymptotic scaling we performed the change of variables suggested by
Parisi \cite{Parisi}, defining  a new temperature $T_E$
proportional to the energy:
\begin{equation}
T_E \;=\; {8N\over N^2-1} \,E\;\;\;,\;\;\;\;\;\beta_E\;=\;{1\over
NT_E}\;\;\;.
\label{betae}
\end{equation}
The ratio of $\Lambda_E$, the $\Lambda$ parameter of the $\beta_E$
scheme, and $\Lambda_L$ is easily obtained from the two loop term
of the energy density:
\begin{equation}
{\Lambda_E\over\Lambda_L}\;=\;\exp \left(\pi {N^2-2\over 4N^2} \right)\;\;\;.
\label{ratioL2}
\end{equation}
Within the $\beta_E$ scheme, the two point function
$\widetilde{G}^E(T_E,p,a)$ must satisfy the renormalization group equation
(neglecting $O\left( a^2\ln^n a\right)$ terms)
\begin{equation}
\left[ -a{\partial \over \partial a} \,+\, \beta_E(T_E){\partial\over \partial
T_E}
\,+\,\gamma_E(T_E)\right] \widetilde{G}_E(T_E,p,a) \;=\;0\;\;.
\label{renormeqE}
\end{equation}
The $\beta$-function of the $\beta_E$ scheme can be written
in the form
\begin{equation}
\beta_E(T_E)\;\equiv\;- a{d\over da}T_E\;=\; {8N^2\over N^2-1} \,C(T)
\,\beta_L(T)
\;\;\;,
\label{betafuncE}
\end{equation}
where
\begin{equation}
C(T)\;=\;{1\over N} {dE\over dT}
\label{spec_heat}
\end{equation}
is the specific heat and $T$ must be considered as a function of $T_E$.
Expanding perturbatively Eq.~(\ref{betafuncE}) and using Eq.~(\ref{WCenergy})
one finds
\begin{equation}
b_{2_E}\;=\;b_{2_L} \,+\, N^2b_0\left( a_2-a_1^2\right) \,+\,
Nb_1a_1\;\;\;.
\label{b2E}
\end{equation}
The scale $\Lambda_E$ is related to the variable
$\beta_E$ by
\begin{equation}
\Lambda_{E} \; = \;
\left( 8\pi\beta_E\right)^{1/2}\,e^{-8\pi\beta_E}
\left[ 1+
{b_1^2-b_0b_{2_E}\over Nb_0^3}\beta_E^{-1} + O\left( \beta_E^{-2}\right)
\right]\;\;\;.
\label{ASYSCE}
\end{equation}
In the large $N$ limit
\begin{equation}
{b_1^2-b_0b_{2_E}\over Nb_0^3}\goto_{N\rightarrow \infty} -0.04009\;\;.
\label{Nlimdle}
\end{equation}

The function $\gamma_E(T_E)$ is easily obtained from the relation
\begin{equation}
\gamma_E(T_E)\;=\;\gamma_L(T)\;\;,
\label{gammaE}
\end{equation}
and therefore
\begin{eqnarray}
\gamma_{2_E}&=&\gamma_{2_L}-\gamma_1Na_1\;=\;{(N^2-1)(N^2-2)\over 128\pi
N^2}\;\;,\nonumber \\
\gamma_{3_E}&=&\gamma_{3_L}\,-\,2\gamma_{2_L}Na_1\,+\,\gamma_1N^2
\left( 2a_1^2-a_2 \right)\;\;.
\label{gamma2E}
\end{eqnarray}

The renormalization equation (\ref{renormeqE})
can be solved following the same procedure used in the standard scheme.
Defining $H_E(T_E,pa)=p^2\widetilde{G}_E(T_E,p,a)$ we obtain
\begin{equation}
H_E(T_E,pa)\;=\; z_E(T_E)^{-1}\;z_E(\Theta_E)\, H(\Theta_E,1)\;\;,
\label{h1e}
\end{equation}
where
\begin{equation}
z_E(T)\;\equiv \;\exp \left( \int dT_E\,{\gamma_E(T_E)\over
\beta_E(T_E)} \right)\;\;,
\label{zfuncE}
\end{equation}
and $\Theta_E\equiv \Theta_{E}(T_E,pa)$ satisfies the equation
\begin{equation}
y{\partial \over \partial y} \Theta_E(T_E,y)\;=\;\beta_E(\Theta_E)\;\;,
\label{Tre}
\end{equation}
whose perturbative solution is obtained from Eq.\ ({\ref{Tr2})
by substituting $b_{2_L}\rightarrow b_{2_E}$ and
$\Lambda_L \rightarrow \Lambda_E$.
The perturbative expansion of $H_E(\Theta_E,1)$ can be found
by reexpressing perturbatively $T$ in terms of $T_E$ in
the r.h.s. of Eq.\ (\ref{lattp}).

Notice that having performed only a coupling redefinition
$T\rightarrow T_E$, it must be
$\widetilde{G}_E(T_E,p,a)=\widetilde{G}_L(T,p,a)$.
But when evaluating  them at some finite order in perturbation
theory we get different results, which are different approximations
of the same quantity.

\section{Numerical results}
\label{numericalresults}

In order to investigate numerically the large $N$ limit of $SU(N)\times
SU(N)$ chiral models, we performed Monte Carlo simulations for several large
values of $N$. We will show numerical results for
$N=6,9,15,21,30$. In Ref.~\cite{chiral1} some results at $N=6,9,15$
were already presented, and in the following we will use some of those data.
A summary of our new large statistics Monte Carlo results is presented in Table
\ref{table}.

In our simulations we used local algorithms containing
overrelaxation procedures.
We implemented the Cabibbo-Marinari algorithm
\cite{Cabibbo} to upgrade $SU(N)$ matrices by updating their $SU(2)$
subgroups.
In most cases, the $SU(2)$ updatings were performed by using the
over-heat-bath algorithm \cite{Petronzio}
(for the ``heat bath'' part of it we used the Kennedy-Pendleton
algorithm \cite{Kennedy}).
A sweep consisted in updating a number of $SU(2)$ subgroups
at all sites of the lattice.
For relatively small values of $N$ ($N\lesssim 6$ say) we chose
to update the $N-1$ diagonal subsequent
$SU(2)$ subgroups of each $SU(N)$ matrix variable.
At larger $N$ we found more efficient to select randomly the
$SU(2)$ subgroups among the ${N(N-1)\over 2}$ subgroups acting on
each $2\times 2$ submatrix.
At each site the $SU(2)$ subgroup identified
by the indices $i,j$ ($1\le i < j \le N$)
was updated with a probability $P={2\alpha\over N-1}$, so that
the average number of $SU(2)$ updatings
per $SU(N)$ site variable was $\bar{n}=\alpha N$.
In our simulations we always chose $\alpha \lesssim 1$,
decreasing $\alpha$ when increasing $N$. At $N=21,30$
we used $\alpha=0.5$.
(We should say that  our choices of the values of $\alpha$
came from a rough study of the performances of the algorithm since the
construction of the ``most'' efficient algorithm was not among our principal
purposes.) Such an algorithm turns out to be quite efficient
in the range of $N$ and $\beta$ (and correlation lengths) we considered.

An important class of observables of the $SU(N)\times SU(N)$ chiral models
can be constructed from the correlation function
$G(x-y)=\langle \;{1\over N} {\rm Re} \;{\rm Tr} \,[ U(x) U(y)^\dagger
]\;\rangle$.

The inverse mass gap $\xi_w$ is extracted from the long distance behavior
of the zero space momentum (wall-wall) correlation function constructed with
$G(x)$.
Moreover we measured the diagonal wall-wall correlation length
$\xi_d$ to test rotation invariance.
$M\equiv 1/\xi_w$ should reproduce in the continuum limit the mass of
the fundamental state.
Another definition of correlation length $\xi_G$ comes from
the second moment of $G(x)$.
In the small momentum regime, $p^2\xi_G^2\ll 1$, we expect the behavior
\begin{equation}
\widetilde{G}(p)\;\simeq\;{Z_G\over M_G^2\,+\,p^2}\;\;,
\label{n1}
\end{equation}
where $\widetilde{G}(p)$ is the Fouries transform of $G(x)$,
$M_G\equiv \xi_G^{-1}$ and $Z_G$ is a constant.
On the lattice we can use the two lowest components of $\widetilde{G}(p)$
to obtain the following definition of $\xi_G$:
\begin{equation}
\xi_G^2 = {1\over4\sin^2\pi/L} \,
\left[{\widetilde G(0,0)\over\widetilde G(0,1)} - 1\right]\;\;\;,
\label{xiG}
\end{equation}
where $\widetilde{G}(k_x,k_y)$ is the lattice Fourier transform of $G(x)$.

In Table~\ref{table}  we present data for the energy density $E$,
the specific heat $C\equiv {1\over N}{dE\over dT}$,
the magnetic susceptibility $\chi\equiv \widetilde{G}(0)$,
the correlation lengths $\xi_G$ and $\xi_w$, the dimensionless ratios
$\xi_G/\xi_w$
and $\xi_d/\xi_w$. Data analysis were performed by using the Jackknife method.
For comparison, in Table \ref{SCtable} we report some strong coupling
results at $N=\infty$ obtained in Sections \ref{strongcoupling}
and \ref{strongcoupling2}.

Finite size effects were carefully checked (see also Ref.~\cite{chiral1}).
Finite size systematic errors in evaluating infinite volume quantities
should be smaller than the statistical errors of all numerical
results presented in this paper.
In all cases we used lattice sizes $L\gtrsim  8.5\xi_G$.

Figs.~\ref{en_N6}, \ref{en_N9} and \ref{en_N15} show the Monte Carlo data of
the energy density and the specific heat
respectively at $N=6$, $N=9$ and $N=15$, with the corresponding strong and weak
coupling series
calculated in the previous sections.
As in other asymptotically free models, at all values of $N$
the specific heat shows a peak, connecting the two different asymptotic
behaviors:
monotonically increasing in the strong coupling region and
decreasing at large $\beta$.
The position of the peak of $C$ should give an estimate
of the strong coupling convergence radius.

As already observed in Ref.~\cite{chiral1} and confirmed by our more recent
data at $N> 15$, increasing $N$ the peak of $C$ moves slightly towards higher
$\beta$ values
and becomes more and more pronounced,
not showing any apparent convergence to a finite value.
This might be an indication of a phase transition at $N=\infty$,
which a rough extrapolation would place at $\beta_{sing}\simeq 0.305$,
with an uncertainty of few per mille.
{}From $N=6$ up to $N=30$, we found the position of the peak
to be very stable with respect to the correlation length:
it occurs at $\xi_G\simeq 2.80$ for $N\ge 6$.

Rotation invariance at distances $d\gtrsim \xi$ is checked
by measuring the ratio $\xi_d/\xi_w$. In Fig.~\ref{xi_diag-side}
the ratio $\xi_d/\xi_w$ is plotted versus $\xi_G$. Data show
effective rotation invariance already for $\xi_G\simeq 2$.
At smaller correlation lengths, discrepancies from one
are well reproduced by the Gaussian model (see Appendix~\ref{BA}), which
predicts
\begin{equation}
{\xi_d\over \xi_w}\;=\; {
\ln \left( {1\over 2\xi_G} + \sqrt{1+{1\over 4\xi_G^2}} \right)
\over
\sqrt{2}\ln \left( {1\over 2\sqrt{2}\xi_G} + \sqrt{1+{1\over 8\xi_G^2}}\right)
}\;\;.
\label{n2}
\end{equation}

Other important tests of scaling are based on the stability of
dimensionless physical quantities, like the ratio $\xi_G/\xi_w$,
which is in general different from one.
Data of $\xi_G/\xi_w$ are shown in Fig.~\ref{xi_G-w}.
Within statistical errors of few per mille, $\xi_G/\xi_w$
is stable for $\xi_G\gtrsim 2$, and independent of $N$ for $N\ge 6$,
showing a rapid convergence to the $N=\infty$ value.
We conclude that in the large $N$ limit
\begin{equation}
{\xi_G\over \xi_w} \;\simeq\; 0.991
\label{n4}
\end{equation}
with an uncertainty of about one per mille.
As shown in Fig.~\ref{xi_G-w},
 the comparison with the Gaussian model prediction
\begin{equation}
{\xi_G\over \xi_w}\;=\;2\xi_G\ln \left( {1\over 2\xi_G}
+\sqrt{1+{1\over 4\xi_G^2}} \,\right)
\label{n3}
\end{equation}
is very good at small correlation lengths $\xi_G\lesssim 2$,
but then discrepancies arise since in the Gaussian case the continuum limit is
one.

Notice that  scaling is observed even around the peak of the specific heat,
even though its behavior
with respect to $N$ suggests the existence of a phase transition at
$N=\infty$. In a sense, at $N=\infty$ the modes responsible for the phase
transition
should be effectively decoupled from those determining the physical
continuum limit, meant
as the renormalization group trajectories where dimensionless
quantities are stable.

Similar considerations  hold for the approach to asymptotic scaling,
when using a ``good'' definition of temperature. A ``good'' definition
of temperature turns out to be $T_E \propto E$ (see Eq.~(\ref{betae})),
whose  corresponding specific heat is, by definition, constant.
{}From the Monte Carlo data of $M\equiv 1/\xi_w$ and the exact result
(\ref{mass-lambda}),
the effective $\Lambda$-parameters $\Lambda_L(N,\beta)$ and
$\Lambda_E(N,\beta_E)$ can be extracted.
Fig.~\ref{LL}  shows the ratio
$\Lambda_L(N,\beta)/\Lambda_{L,2l}(N,\beta)$, where
$\Lambda_{L,2l}(N,\beta)$ is the corresponding two loop function:
$\Lambda_{L,2l}(N,\beta)= \left( 8\pi\beta\right)^{1/2}\,e^{-8\pi\beta}$.
Around the region where the specific heat has a peak,
also $\Lambda_L(N,\beta)/\Lambda_{L,2l}(N,\beta)$ shows a peak,
whose shape tends to be singular when $N\rightarrow\infty$ for
$\beta_{sing}\simeq 0.305$,
consistently with the extrapolation from the specific heat.
The observed peaks in the ratio $\Lambda_L(N,\beta)/\Lambda_{L,2l}(N,\beta)$
correspond to dips in the $\beta$-functions,
and the singularity of $\Lambda_L(\infty,\beta)$ would reveal
a singularity of the $N=\infty$ $\beta$-function.
The similarity with the behavior of the specific heat suggests
that the peak of $C$ and the dip of the $\beta$-function have the same origin.
They are presumably related to complex $\beta$-singularities of the partition
function
close to the real axis \cite{Marinari}, that for $N\rightarrow\infty$
should finally pinch the real axis.

As already observed in other contexts (see for example
\cite{Karsch,Wolff,amsterdam}),
the approach to asymptotic scaling gets an impressive improvement in the
$\beta_E$
scheme, where the dip of the $\beta$-function disappears at all values of $N$.
Fig.~\ref{LE} shows the ratio
$\Lambda_E(N,\beta_E)/\Lambda_{E,2l}(N,\beta_E)$,
where $\Lambda_E(N,\beta_E)$ is the effective $\Lambda$-parameter of the
$\beta_E$ scheme, and $\Lambda_{E,2l}(N,\beta_E)$ is the corresponding two
loop function.
Unlike $\Lambda_L(N,\beta)$, $\Lambda_E(N,\beta_E)$
appears to approach a smooth function
$\Lambda_E(\infty,\beta_E)$, that is well approximated by the two
loop formula $\Lambda_{E}(\infty,\beta_E)\simeq \left(
8\pi\beta_E\right)^{1/2}\,e^{-8\pi\beta_E}$
even around the peak of the specific heat, which is placed  at $\beta_E\simeq
0.220$ for $N\ge 6$ (the peak position being much more stable in $N$
when described by the variable $\beta_E$ instead of $\beta$).

In Section~\ref{weakcoupling} we found that in
the standard and the energy schemes
the linear corrections to the two loop lattice
scale in Eqs.~(\ref{ASYSC}) and (\ref{ASYSCE}) are small
and of the same order of
magnitude, although of opposite sign.
Therefore perturbative arguments do not
explain the failure of the standard and the
success of the $\beta_E$ scheme with respect to achieving asymptotic scaling.
We believe the flattening of the specific heat to be the key feature of the
$\beta_E$ scheme. A coupling transformation
eliminating the peak of the specific heat should move the complex
$\beta$-singularities
away from the real axis, and therefore improve the approach to asymptotic
scaling.

The above considerations are consistent with the relationship
(\ref{betafuncE}).
Indeed, assuming that at $N=\infty$ the $\beta$-function of the $\beta_E$
scheme is not
singular (as data seem to show) and the specific heat
has a divergence at $\beta_{sing}$, then
Eq.~(\ref{betafuncE}) predicts a singularity
in the $N=\infty$ $\beta$-function of the standard scheme at $\beta_{sing}$.

Asymptotic scaling within the $\beta_E$ scheme can be also checked
by using the strong coupling estimates of $E$, $\xi_w$ and $\xi_G$.
In Fig.~\ref{asySC} we show the results using the strong coupling
series obtained in Sections~\ref{strongcoupling} and \ref{strongcoupling2}
($\;E$ up to $O\left( \beta^{14}\right)$,
$\xi_w$ up to $O\left( \beta^{8}\right)$ and
$\xi_G$ up to $O\left( \beta^{9}\right)\;$)
at $N=\infty$, and for $\beta_E < 0.220$, which should be the  approximate
position of the specific heat singularity.
The comparison with the predicted mass-$\Lambda$ parameter ratio is very
good, especially for $\xi_G$ whose series has alternate signs.

Another quantity, we found to be very sensitive to the change of variable
$T\rightarrow T_E$, is the renormalization constant $Z_G$ introduced in
Eq.~(\ref{n1}). Its dependence on $\beta$ can be determined by
renormalization group considerations, indeed it must satisfy
Eq.~(\ref{gamma}). One finds
\begin{eqnarray}
Z_G&=&c(N)\,N^{\gamma_1/b_0} \,z(T)^{-1}\;=\;c(N)\,\beta^{-\gamma_1/b_0}\,
\left[ 1\,+\,O\left( {1\over \beta}\right)\right]\nonumber \\
&=&c(N)\,N^{\gamma_1/b_0} \,z_E(T_E)^{-1}\;=\;c(N)\,\beta_E^{-\gamma_1/b_0}\,
\left[ 1\,+\,O\left( {1\over \beta_E}\right)\right]\;\;,
\label{n5}
\end{eqnarray}
where the function $z(T)$ and $z_E(T_E)$ were defined in
Eqs.~(\ref{zfunction}) and (\ref{zfuncE}),
and $c(N)$ is a constant independent of the regularization scheme.
In Figs.~\ref{z_b} and \ref{z_be} we plotted respectively the quantities
$c_L(N,\beta)\equiv Z_G\beta^{\gamma_1/b_0}$ and $c_E(N,\beta_E)\equiv
Z_G\beta_E^{\gamma_1/b_0}$.
Fig.~\ref{z_b} shows also the strong coupling estimate of
$c_L(\infty,\beta)$ obtained from Eq.~(\ref{m27}).
$c_L(N,\beta)$ appears to be far from being approximately constant.
It tends to be singular when $N\rightarrow\infty$ at $\beta_{sing}\simeq
0.305$,
consistently with the previous estimates of the singularity.
On the contrary, $c_E(N,\beta_E)$ shows a much smoother behavior and a rapid
convergence at large $N$.
Again, perturbative arguments do not help to explain such different
behaviors, since perturbative corrections are small for both schemes.
Indeed from Eqs.~(\ref{zfunction}) and (\ref{zfuncE}) we can derive
the following relationships at $N=\infty$:
\begin{equation}
c_L(\infty,\beta)\;=\;c(\infty)\,\left[
1\,+\,{0.08521\over \beta}\,+\,{0.01314\over \beta^2}\,+ \,
O\left({1\over\beta^3}\right) \right]\;\;,
\label{n6}
\end{equation}
and
\begin{equation}
c_E(\infty,\beta_E)\;=\;c(\infty)\,\left[
1\,+\,{0.02271\over \beta_E}\,+\,{0.00079\over \beta_E^2}\,+ \,
O\left({1\over\beta_E^3}\right) \right]\;\;.
\label{n7}
\end{equation}
We could get an estimate of  $c(\infty)$ from the data at the largest available
$\beta_E$ values,
obtaining $c(\infty)\simeq 0.10$
(the perturbative corrections in Eq.~(\ref{n7}) are about $10\%$ at
$\beta_E\simeq 0.25$).

In Section \ref{strongcoupling2} we found that at large $N$ the correlation
function
breaks the Gaussian form at relatively high orders in the strong coupling
expansion.
Therefore in the very strong coupling domain, $\widetilde{G}(k)$ should be
well approximated by Eq.~(\ref{n1}) everywhere and not only for
$k^2\ll M_G^2$.
When the Gaussian approximation works we should find
\begin{equation}
M(k)\;\equiv\;{\widetilde{G}(k)\over \widetilde{G}(0)}
\left( M_G^2\,+\,\hat{k}^2\right)\;\simeq\;1\;\;.
\label{n8}
\end{equation}
Fig. \ref{gautest} shows the components $(k,0)$ and $(k,k)$ of $M(k_1,k_2)$
at $N=30$ and  for various values of $\beta$ chosen inside the expected
convergence region of the strong coupling expansion.
As expected from the strong coupling considerations, at $\beta=0.20$
the propagator is effectively Gaussian. Discrepancies, although small,
are clear for $\beta\ge 0.28$, especially in the diagonal components.

In order to study the continuum limit of $\widetilde{G}(k)$,
we consider the dimensionless function
\begin{equation}
L(k;\beta,N)\;\equiv\; {\widetilde{G}(0;\beta,N)\over\widetilde{G}(k;\beta,N)}
\,-\,1\;\;.
\label{n9}
\end{equation}
$L(k;\beta,N)$ is renormalization group invariant and therefore for
$k^2a^2\lesssim 1$
\begin{equation}
L(k;\beta,N)\;\simeq\;\bar{L}(k^2\xi_G^2,N)\;\;,
\label{n10}
\end{equation}
the function $\bar{L}(y,N)$ being independent of $\beta$.
Notice that in the Gaussian model we would have $\bar{L}(y)=y$.
In Figs.~\ref{Lk6} and \ref{Lk9}
we plotted the components $(k,0)$ and $(k,k)$ of
$L(k_1,k_2;\beta,N)$ versus $k^2\xi_G^2$ obtained respectively at $N=6$
for $\beta=0.31$ and $\beta=0.33$, and
at $N=9$  for $\beta=0.31$ and
$\beta=0.32$. All sets of data follow  a single curve for
$k^2\xi_G^2\lesssim \xi_G^2 /a^{2}$, which  must be the continuum function
$\bar{L}(y,N)$ at the given $N$.
Then discrepancies, i.e. scaling violations, arise.
Comparing data at different $\beta$, we also learn that:
(i) When there is rotation invariance at a scale $p$, i.e.
$L(p,0;\beta,N)\simeq L(p/\sqrt{2},p/\sqrt{2};\beta,N)$, then
$L(p;\beta,N)\simeq\bar{L}(p^2\xi_G^2,N)$.
(ii) The diagonal components $L(k,k;\beta,N)$ are closer to the continuum
limit than the $L(k,0;\beta,N)$ ones.

Fig. \ref{LkN} gathers data of $L(k,k;\beta,N)$ at different values of $N$ and
$\beta$. At relatively low values of $k^2\xi_G^2$, all sets of data follow
approximatively a single curve, showing that $\bar{L}(y,N)$ rapidly converges
to its large $N$ limit. $N=6,9$ data give already a good approximation
of $\bar{L}(y;\infty)$.
At larger $k^2\xi_G^2$, the scattering of the curves for different $N$
is essentially due to scaling violations,
because we used data at different $\xi_G$.

The extrapolation of  $\bar{L}(y,N)$ to negative $y$ should provide
another estimate of the ratio $M/M_G\equiv \xi_G/\xi_w$.
The solution $y_0$ of the mass gap equation $\bar{L}(y)+1=0$
is  $y_0=-M^2/M_G^2$.
At sufficiently small $y$, $\bar{L}(y,\infty)$ shows an approximate
Gaussian behavior, indeed for $y\lesssim 5$ it is well
reproduced by a polynomial function
$y+by^2$ with $b$ very small: $b\simeq 0.01$.
By using this polynomial function to extrapolate, we find $y_0\simeq -1$
within 1\%, consistently with the more precise estimate (\ref{n4}).

 We finally compare the Monte Carlo data of $\widetilde{G}(k)$ with the
perturbative solutions $\widetilde{G}(k)_p$
of the renormalization group equations (\ref{h1}) and (\ref{h1e}).
In Figs. \ref{GN6} and \ref{GN9} we show data
respectively at $N=6$ ($\beta=0.33$) and $N=9$ ($\beta=0.32$),
with their corresponding renormalization group equation solutions at the lowest
order and
at the next two orders,
calculated in the standard and the $\beta_E$ schemes (see
Section~\ref{weakcoupling}).
In particular the lowest order solution of the renormalization group
equation for $\widetilde{G}(k)$ is
\begin{equation}
\widetilde{G}(k)_{p,1}\;=\;\left( {N^2-1\over 2N}\right)
{T^{\gamma_1/b_0}\over k^2}
\left( {b_0\over 2} \ln {k^2\over \Lambda_L^2 }\right)^{\gamma_1/b_0-1}
\label{n11}
\end{equation}
in the standard scheme, and
\begin{equation}
\widetilde{G}(k)_{p,1}\;=\;\left( {N^2-1\over 2N}\right)
{T_E^{\gamma_1/b_0}\over k^2}
\left( {b_0\over 2} \ln {k^2\over\Lambda_E^2} \right)^{\gamma_1/b_0-1}
\label{n12}
\end{equation}
in the $\beta_E$ scheme.
In $\widetilde{G}(k)_p$ the scale is provided by the relationship
\begin{equation}
\Lambda_{L,E}\;=\;\left( {\Lambda_{L,E} \over M} \right)\times M
\;=\;{M\over R_{L,E}}\;\;,
\label{n13}
\end{equation}
where $R$ is the mass-$\Lambda$ parameter ratio in the corresponding scheme,
and $M\equiv 1/\xi_w$ is taken from the Monte Carlo simulation.

Again, the best result comes from the $\beta_E$ scheme, and in
particular from its lowest order approximation.
The $\Lambda_E$ parameter shows a similar behavior too,
its comparison with the corresponding
two loop formula is already very good, and
the $O\left( \beta_E^{-1}\right)$ correction does not improve it.
This seems to be a general feature of the $\beta_E$ scheme:
its lowest order renormalization group estimates do not
need $O\left( \beta_E^{-1}\right)$ corrections, since they are
already  very good. On the other hand, we should
take into account that this scheme comes from a non perturbative change
of variable.

According to the conjectured $S$-matrix \cite{Abdalla,Wiegmann1,Wiegmann2}
and the large $N$ factorization, 2-d $SU(N)$ chiral theories should
describe free particles in the limit $N\rightarrow \infty$.
But when considering the two point Green's function $G(x)$,
we see that the realization of such physical properties is not trivial at all.
Indeed, renormalization group considerations (see
Section~\ref{weakcoupling}) tell us that
in the limit $N\rightarrow\infty$ the asymptotic behavior of $G(x)$ for $x\ll
1/M$ is
\begin{equation}
G(x)\;\sim\;\ln^2\left( {1\over x M}\right) \;\; ,
\label{n14}
\end{equation}
while a free Gaussian Green's function behaves as $\ln\left( 1/x \right)$.
Then $G(x)$, at small $x$,  seems to describe the propagation
of a composite object formed by two elementary Gaussian
excitations, suggesting an interesting  hadronization picture:

In the large $N$ limit, the Lagrangian fields $U$, playing the role of
non-interacting hadrons, are constituted by two confined particles, which
appear free in the large momentum limit, due to asymptotic freedom.

\appendix{}
\label{AA}

For low-dimensional
representations the $\widetilde{z}_{(r)}$ are (known) monomials in $\beta$,
while their counterparts $\widetilde{z}_{(r+1^N)}$ may be reconstructed
starting from the quantities
\begin{equation}
\Gamma_{l_i;m_j}\;=\;\langle \;{\rm det}\,U \,\prod_{i=1}^p {\rm Tr} U^{l_i}\,
\prod_{j=1}^q {\rm Tr} U^{\dagger\,m_j} \;\rangle\;\;,
\label{s24}
\end{equation}
which in turn are recursively constructed by applying differential
operators and algebraic Schwinger-Dyson equations (Ref.~\cite{Green},
Appendix~D)
to $\Gamma_0$
\begin{equation}
\Gamma_0\;=\; \widetilde{z}_{(1^N)}\;=\;J_N(2N\beta)\,+\,O\left(
\beta^{3N+2}\right)\;\;.
\label{s25}
\end{equation}
We have found the general form of a few interesting classes of character
coefficients:
\begin{equation}
d_{(l;0)}\widetilde{z}_{(l;1^N)}\;\simeq \;
\sum_{k=0}^l {(N\beta)^{l-k}\over (l-k)!} (-1)^k J_{N+k}\;\;,
\label{s26}
\end{equation}
\begin{equation}
d_{(1^l;0)} z_{(1^l;1^N)}\;\simeq \;
{(N\beta)^l\over l!} J_N \,-\, {(N\beta)^{l-1}\over (l-1)!} J_{N+1} \;\;,
\label{s27}
\end{equation}
\begin{equation}
d_{(0;1^l)} z_{(0;1^l+1^N)}\;\simeq \;
\sum_{k=0}^l {(N\beta)^{l-k}\over (l-k)!} J_{N-k}\;\;,
\label{s28}
\end{equation}
\begin{equation}
d_{(0;l)} z_{(0;l+1^N)}\;\simeq \;
{(N\beta)^l\over l!} J_N \,+\, {(N\beta)^{l-1}\over (l-1)!} J_{N-1} \;\;.
\label{s29}
\end{equation}
Moreover we have explicitly computed the $SU(N)$ character coefficients for
the first few low-dimensionality representations
\begin{equation}
d_{(2;0)}z_{(2;0)}\;\simeq\;{1\over 2}(N\beta)^2\,+\,J_{N+2} \,+\,
2N\beta J^\prime_N\;\;,
\label{s30}
\end{equation}
\begin{equation}
d_{(1^2;0)}z_{(1^2;0)}\;\simeq\;{1\over 2}(N\beta)^2\,+\,J_{N-2} \,+\,
2N\beta J^\prime_N\;\;,
\label{s31}
\end{equation}
\begin{equation}
d_{(1;1)}z_{(1;1)}\;\simeq\;(N\beta)^2\,-\,2J_{N} \,+\,
4N\beta J^\prime_N\;\;,
\label{s32}
\end{equation}
\begin{equation}
d_{(3;0)}z_{(3;0)}\;\simeq\;{1\over 6}(N\beta)^3\,+\,
{1\over 2}(N\beta)^2J_{N-1}\,-\,
{1\over 2}(N\beta)^2J_{N+1}\,+\,
N\beta J_{N+2}\,-\,J_{N+3}\;\;,
\label{s33}
\end{equation}
\begin{equation}
d_{(2,1;0)}z_{(2,1;0)}\;\simeq\;{1\over 3}(N\beta)^3\,+\,
N\beta J_{N-2}\,+\,(N\beta)^2 J_{N+1}\,-\,
(N\beta)^2 J_{N+1}\,+\,N\beta J_{N+3}\;\;,
\label{s34}
\end{equation}
\begin{equation}
d_{(1^3;0)}z_{(1^3;0)}\;\simeq\;{1\over 6}(N\beta)^3\,+\,
J_{N-3}\,+\,N\beta J_{N-2}\,+\,
{1\over 2}(N\beta)^2 J_{N-1}\,-\,{1\over 2}(N\beta)^2 J_{N+1}\;\;,
\label{s35}
\end{equation}
\begin{equation}
d_{(2;1)}z_{(2;1)}\;\simeq\;{1\over 2}(N\beta)^3\,+\,
{3\over 2}(N\beta)^2 J_{N-1}\,-\,2N\beta J_{N}\,+\,
\left[ 1-{3\over 2}(N\beta)^2\right]
J_{N+1}\,-\,N\beta J_{N+2}\;\;,
\label{s36}
\end{equation}
\begin{equation}
d_{(1^2;1)}z_{(1^2;1)}\;\simeq\;{1\over 2}(N\beta)^3\,+\,
N\beta J_{N-2}\,-\,
\left[ 1-{3\over 2}(N\beta)^2\right]
J_{N-1}\,-\,2N\beta J_{N}\,-\,
{3\over 2}(N\beta)^2 J_{N+1}\;\;.
\label{s37}
\end{equation}

\appendix{}
\label{BA}

In order to improve our understanding, we briefly considered the Gaussian
model, which bears a wide resemblance to the very strong
coupling behavior of the chiral models for sufficiently large $N$.
The Gaussian model is exactly solvable, and one finds that, for arbitrary
values of the coupling and arbitrary direction $\theta$ in the spatial
lattice, Eq.~(\ref{u4}) becomes \cite{Muller,Campo-Rossi-rev}
\begin{equation}
\hat{p}^2\,+\,m^2\;=\;0\;\;,\;\;\;\;\;\;\;\hat{p}^2=4\sum_\mu \sin^2
\left( {p_\mu\over 2}\right)\;\;,
\label{u7}
\end{equation}
and is solved by
\begin{equation}
\mu(\theta )\;=\; \cos\theta \,{\rm arsh} \left(\nu\cos\theta\right) \,+\,
\sin\theta \,{\rm arsh} \left(\nu\sin\theta\right)\;\;,
\label{u8}
\end{equation}
where
\begin{equation}
\nu (\theta)\;=\;m\sqrt{2+{m^2\over 4}} \left[
1 \,+\, \sqrt{1-m^2(8+m^2)\left( {\cos2\theta\over 4+m^2}\right)^2}\;
\right]^{-1/2}\;\;.
\label{u9}
\end{equation}
Introducing the auxiliary strong coupling variable
\begin{equation}
\gamma\;=\;{1\over 4+m^2}
\label{u10}
\end{equation}
(related to $\beta$ by $\beta=\gamma + O(\gamma^3)$, odd function
of $\gamma$),
we obtain
\begin{equation}
\nu (\theta) \;=\; {1\over 2\gamma} {\sqrt{1-16\gamma^2}\over
\sqrt{1+\sqrt{\sin^2(2\theta)+16\gamma^2\cos^2(2\theta)}}}\;\;,
\label{u11}
\end{equation}
that is, $\nu$ is an odd function of $\gamma$ for all $\theta \neq 0$.
The crucial observation however concerns the limit $\theta\rightarrow 0$,
where
\begin{equation}
\nu(0)\;=\;{1\over 2\gamma}\sqrt{1-4\gamma}
\label{u12}
\end{equation}
which is not a function of $\gamma$ with definite parity.
Therefore, while for all $\theta \neq 0$ the quantity
\begin{equation}
\rho(\theta)\;=\; \mu(\theta) \,+\, (\cos\theta +\sin\theta)\ln
\beta(\gamma)
\label{u13}
\end{equation}
is an even function of $\gamma$, this property does not hold at $\theta=0$.
When we consider Green's functions, we recognize that in strong coupling
\begin{equation}
G(x_1,x_2)\;=\;\beta^{x_1+x_2}H(x_1,x_2;\beta)
\label{u14}
\end{equation}
where $H$ is always an even function of $\beta$ (and $\gamma$).
As a consequence, the purely kinematical singularity at $\theta=0$ in
Eq.~(\ref{u11}) prevents the general relationship
\begin{equation}
\rho(\theta)\;=\;
-\,\lim_{|x|\rightarrow \infty} {\ln H(x,\beta)\over |x|}\;\;
\label{u15}
\end{equation}
from holding at $\theta=0$.
For all other orientations exponentiation can be shown to hold,
and in particular the principal diagonal correlation (a quantity of
fundamental relevance for tests of rotation invariance) has the proper
exponential decay and the corresponding mass gap $\mu_d$ can be shown to
coincide with the value extracted from the diagonal wall-wall correlation:
\begin{eqnarray}
&&\widetilde{G}^{-1}(p_1={i\mu_d\over\sqrt{2}},\;p_2={i\mu_d\over\sqrt{2}})\;=\;
0\nonumber \\
&&\mu_d\;=\;
-\,\lim_{|x|\rightarrow \infty} {\ln G_{diag}(x_1=x_2=|x|/\sqrt{2})
\over |x|}\;\;.
\label{u16}
\end{eqnarray}
Notice that for arbitrary $\theta$ the two definitions
do not strictly coincide.

\appendix{}
\label{BB}

We could establish the following recursive relationships
\begin{equation}
C_0(x_1,x_2)\,-\,C_0(x_1-1,x_2)\,-\,C_0(x_1,x_2-1)\;=\;0\;\;,
\label{u19}
\end{equation}
with boundary conditions
\begin{equation}
C_0(x_1,0)\;=\;C_0(0,x_2)\;=\;1
\label{u20}
\end{equation}
and (well known) solution
\begin{equation}
C_0(x_1,x_2)\;=\;
\left( \begin{array}{c} x_1+x_2\\ x_1\end{array}\right)
\;\equiv\;\left( \begin{array}{c} x_1+x_2\\ x_2\end{array}\right)\;\;;
\label{uu21}
\end{equation}
\begin{eqnarray}
&&C_2(x_1,x_2)\,-\,C_2(x_1-1,x_2)\,-\,C_2(x_1,x_2-1)\;=\nonumber \\
&&C_0(x_1-1,x_2+1)\,+\,C_0(x_1+1,x_2-1)\,-\,C_0(x_1-1,x_2)
\,-\,C_0(x_1,x_2-1)\;\;,
\label{u22}
\end{eqnarray}
with boundary conditions
\begin{eqnarray}
C_2(x_1,0)&=& x_1(x_1+1)\;\;,\nonumber \\
C_2(0,x_2)&=& x_2(x_2+1)\;\;,
\label{u23}
\end{eqnarray}
and solution
\begin{equation}
C_2(x_1,x_2)\;=\;
\left( \begin{array}{c} x_1+x_2\\ x_1\end{array}\right)\,
\left[ {x_1(x_1+1)\over x_2+1}\,+\,{x_2(x_2+1)\over x_1+1}\right]\;\;;
\label{uu24}
\end{equation}
\begin{eqnarray}
&&C_4(x_1,x_2)\,-\,C_4(x_1-1,x_2)\,-\,C_4(x_1,x_2-1)\;=\;
\nonumber \\
&&C_2(x_1+1,x_2-1)\,+\,C_2(x_1-1,x_2+1)
\,-\,C_2(x_1,x_2)\,+C_0(x_1+2,x_2-1)\nonumber \\
&&\;+\,C_0(x_1-1,x_2+2)\,+\,C_0(x_1,x_2-2)
\,+\,C_0(x_1-2,x_2)\,-\,3C_0(x_1,x_2)\;\;,
\label{u25}
\end{eqnarray}
with boundary conditions
\begin{eqnarray}
C_4(x_1,0)&=&{x_1^2(x_1+1)^2\over 4}\,+\,{x_1(x_1+1)\over
2}\,+\,4x_1\;\;,\nonumber \\
C_4(0,x_2)&=&{x_2^2(x_2+1)^2\over 4}\,+\,{x_2(x_2+1)\over
2}\,+\,4x_2\;\;,
\label{u26}
\end{eqnarray}
and solution
\begin{eqnarray}
C_4(x_1,x_2)\;=&&
\left( \begin{array}{c} x_1+x_2\\ x_1\end{array}\right)\;
\Biggl[ x_1x_2 + 2x_1+2x_2+
{x_2(x_2+3)\over x_1+1}\,+\,{x_1(x_1+3)\over x_2+1}\nonumber \\
&&\;\;\;+{(x_2-1)x_2(x_2+1)(x_2+2)\over 2(x_1+1)(x_1+2)}+
{(x_1-1)x_1(x_1+1)(x_1+2)\over 2(x_2+1)(x_2+2)}-
{2x_1x_2\over x_1+x_2}\Biggr]\;\;.
\label{uu27}
\end{eqnarray}

The function $B$ can be computed by the relationship
\begin{equation}
B(x_1,x_2)\;=\;\sum_a a \,C_0^{(a)}(x_1,x_2)\;\;,
\label{u30}
\end{equation}
where $C_0^{(a)}(x_1,x_2)$ is the number of minimal self-avoiding
paths connecting the origin with $x$ and forming $a$ right angles.
$C_0^{(a)}$ satisfies the normalization condition
\begin{equation}
\sum_a C_0^{(a)}(x_1,x_2)\;=\; C_0(x_1,x_2)\;\;.
\label{u31}
\end{equation}
Solving appropriate recursive equations it is possible to prove that
\begin{eqnarray}
\sum_{x_1,x_2} C_0^{(a)}(x_1,x_2) t_1^{x_1} t_2^{x_2}&=&
\left( {t_1\over 1-t_1}+{t_2\over 1-t_2}\right)
\left({t_1 t_2\over (1-t_1)(1-t_2)}\right)^{a\over 2}\;\;\;\;\;\;\;{\rm
for} \;\;\;{\rm even} \;\;a\;\;,\nonumber \\
\sum_{x_1,x_2} C_0^{(a)}(x_1,x_2) t_1^{x_1} t_2^{x_2}&=&
2\left({t_1 t_2\over (1-t_1)(1-t_2)}\right)^{a+1\over 2}\;\;\;\;\;\;\;{\rm
for}\;\;\;{\rm  odd}\;\;a\;\;,
\label{u32}
\end{eqnarray}
and as a consequence
\begin{equation}
\sum_{x_1,x_2} B(x_1,x_2) t_1^{x_1} t_2^{x_2}\;=\; {2t_1t_2\over
(1-t_1-t_2)^2}\;\;.
\label{u33}
\end{equation}
We then trivially obtain
\begin{equation}
B(x_1,x_2)\;=\; {2x_1x_2\over x_1+x_2}\,C_0(x_1,x_2)\;\;.
\label{uu34}
\end{equation}

\appendix{}
\label{BC}

We report here the $N=\infty$ strong coupling series of
some relevant  quantities:

\begin{equation}
E\;=\;1-\beta-2\beta^3-6\beta^5-38\beta^7-240\beta^9-
1812\beta^{11}-14126\beta^{13}
+O\left(\beta^{15}\right)\;,
\label{bc1}
\end{equation}

\begin{eqnarray}
\chi\;=&&1+4\beta+12\beta^2+36\beta^3+100\beta^4+284\beta^5+796\beta^6
+2276\beta^7\nonumber \\
&&+6444\beta^8+18572\beta^9+53284\beta^{10}
+O\left(\beta^{11}\right)\;,
\label{bc2}
\end{eqnarray}

\begin{equation}
Z_G\;=\;\beta^{-1}\left[ 1-\beta^2-6\beta^4-4\beta^5-20\beta^6
-24\beta^7-148\beta^8-240\beta^9+
O\left(\beta^{10}\right)\right]\;,
\label{bc3}
\end{equation}

\begin{equation}
M_{G}^2\;=\;
\beta^{-1}\left[ 1-4\beta+3\beta^2-2\beta^4-4\beta^5-12\beta^6
-40\beta^7+84\beta^8-320\beta^9+
O\left(\beta^{10}\right)\right]\;,
\label{bc4}
\end{equation}

\begin{equation}
\mu_{side}\;=\;-\ln \beta -2\beta -{2\over 3}\beta^3 -2\beta^4
-{42\over 5}\beta^5 -8\beta^6 -{310\over 7}\beta^7 - 84\beta^8
+O\left(\beta^{9}\right)\;,
\label{bc5}
\end{equation}

\begin{equation}
\mu_{diag}\;=\;\sqrt{2} \left[ -\ln 2\beta - \beta^2 -3\beta^4
-{119\over 6}\beta^6
+O\left(\beta^{8}\right)\right]\;.
\label{bc6}
\end{equation}



\figure{
Energy and specific heat versus $\beta$ for $N=6$.
The full and dashed lines
represent respectively the strong coupling
(up to $O\left( \beta^{12}\right)$ for $E$ and
up to $O\left( \beta^{13}\right)$ for $C$)
and the weak coupling series.
\label{en_N6}}

\figure{
Energy and specific heat versus $\beta$ for $N=9$.
The full and dashed lines
represent respectively the strong coupling
(up to $O\left( \beta^{14}\right)$ for $E$ and
up to $O\left( \beta^{15}\right)$ for $C$)
and the weak coupling series.
\label{en_N9}}

\figure{
Energy and specific heat versus $\beta$ for $N=15$.
The full and dashed lines
represent respectively the strong coupling
(up to $O\left( \beta^{14}\right)$ for $E$ and
up to $O\left( \beta^{15}\right)$ for $C$)
and the weak coupling series.
\label{en_N15}}

\figure{
The ratio $\xi_d/\xi_w$ versus $\xi_G$.
The full line represents the Gaussian prediction
(\ref{n2}).
\label{xi_diag-side}}

\figure{
The ratio $\xi_G/\xi_w$ versus $\xi_G$.
The full line represents the Gaussian prediction
(\ref{n3}).
The dashed line  is the result of a fit.
\label{xi_G-w}}

\figure{$\Lambda_L(N,\beta)/\Lambda_{L,2l}(N,\beta)$ versus $\beta$.
The dotted lines connecting
different sets of data are drawn to guide the eyes.
\label{LL}}

\figure{$\Lambda_E(N,\beta_E)/\Lambda_{E,2l}(N,\beta_E)$ versus $\beta_E$.
\label{LE}}

\figure{Asymptotic scaling test by using strong coupling estimates.
The dotted line represents the exact result (\ref{mass-lambda}).
\label{asySC}}

\figure{$c_L=Z_G\beta^{\gamma_1/b_0}$ versus $\beta$.
The dashed line shows the strong coupling series
for $N=\infty$.
The dotted lines connecting
different sets of data are drawn to guide the eyes.
\label{z_b}}

\figure{$c_E=Z_G\beta_E^{\gamma_1/b_0}$ versus $\beta_E$.
\label{z_be}}

\figure{$M(k,0)$ and $M(k,k)$ versus $\hat{k}^2$ at $N=30$.
\label{gautest}}

\figure{$L(k,0)$ and $L(k,k)$ versus $k^2\xi_G^2$ at $N=6$
and for $\beta=0.31$ and $\beta=0.33$.
The dotted lines connecting
different sets of data are drawn to guide the eyes.
The dashed line represents the Gaussian prediction.
\label{Lk6}}

\figure{$L(k,0)$ and $L(k,k)$ versus $k^2\xi_G^2$ at $N=9$
and for $\beta=0.31$ and $\beta=0.32$. The dotted lines connecting
different sets of data are drawn to guide the eyes.
The dashed line represents the Gaussian prediction.
\label{Lk9}}

\figure{$L(k,k)$ versus $k^2\xi_G^2$.
\label{LkN}}

\figure{$G(k,0)$ and $G(k,k)$ versus $k^2\xi_G^2$ at $N=6$ and
$\beta=0.33$.
The low and high sets of lines (full, dashed and dotted)
show respectively the renormalization group predictions of
the standard and the $\beta_E$ schemes.
Full, dashed and dotted lines represent respectively
the lowest, and the one and two next order perturbative solutions.
\label{GN6}}

\figure{$G(k,0)$ and $G(k,k)$ versus $k^2\xi_G^2$ at $N=9$ and
$\beta=0.32$.
The low and high sets of lines (full, dashed and dotted)
show respectively the renormalization group predictions of
the standard and the $\beta_E$ schemes.
Full, dashed and dotted lines represent respectively
the lowest, and the one and two next order perturbative solutions.
\label{GN9}}


\begin{table}
\squeezetable
\caption{Summary of the numerical results. Errors of data marked by an
asterisk  could be underestimated. When more than one lattice size appear,
the corresponding results were obtained collecting
data of simulations at the reported lattice sizes
(which were, in all cases, in agreement within the errors).}
\label{table}
\begin{tabular}{r@{}lrcr@{}lr@{}lr@{}lr@{}lr@{}lr@{}lr@{}lr@{}l}
\multicolumn{2}{c}{$\beta$}&
\multicolumn{1}{r}{$N$}&
\multicolumn{1}{c}{$L$}&
\multicolumn{2}{c}{$E$}&
\multicolumn{2}{c}{$C$}&
\multicolumn{2}{c}{$\chi$}&
\multicolumn{2}{c}{$\xi_G$}&
\multicolumn{2}{c}{$\xi_w$}&
\multicolumn{2}{c}{$\xi_G/\xi_w$} &
\multicolumn{2}{c}{$\xi_d/\xi_w$} \\
\tableline
0&.20 & 15 &18,21 &0&.781405(7) &0&.0527(3) &2&.9380(7) & 0&.786(3) &
0&.8360(4) &0&.941(3)  & 0&.9724(4)\\
0&.20 & 21 &18,21 &0&.781427(7) &0&.0527(3) &2&.9381(5) & 0&.788(2) &
0&.8358(3) &0&.943(2)  & 0&.9726(4)\\
0&.20 & 30 &18    &0&.781422(10)&0&.0522(6) &2&.9379(8) & 0&.787(2) & 0&.834(2)
 &0&.944(3)  & 0&.973(2)\\\hline
0&.28 & 15 &18,24 &0&.65000(5)  &0&.191(5) &7&.249(6)  & 1&.560(4) & 1&.587(2)
&0&.9834(13)& 0&.9922(6)\\
0&.28 & 21 &18,24 &0&.65290(3)  &0&.170(3) &7&.069(5)  & 1&.532(3) &
1&.5605(10)&0&.9819(11)& 0&.9917(4)\\
0&.28 & 30 &18    &0&.65352(3)  &0&.163(4) &7&.032(5)  & 1&.529(3) &
1&.5544(15)&0&.9837(12)& 0&.9916(5)\\\hline
0&.29 &  9 &24,30 &0&.58772(7)  &0&.412(5) &13&.32(3)  & 2&.369(10)& 2&.395(9)
&0&.9889(9) & 1&.000(2)\\
0&.29 & 30 &21    &0&.63058(3)  &0&.208(6) & 8&.497(8) & 1&.740(4) & 1&.765(5)
&0&.9857(7) & 0&.992(2)\\\hline
0&.295&  9 &24,30,36,42 &0&.56278(6)  &0&.444(5) &18&.03(4)  & 2&.913(9) &
2&.949(13) &0&.988(2)  & 0&.996(3)\\\hline
0&.30 & 15 &24,30,36,42 &0&.56806(4)  &0&.68(2)  &16&.574(13)& 2&.742(5) &
2&.767(9)  &0&.9907(10)& 1&.000(1)\\
0&.30 & 21 &24    &0&.58799(14) &0&.65(3)  &12&.91(3)  & 2&.309(7) & 2&.333(7)
&0&.9899(7) & 0&.997(2)\\
0&.30 & 30 &24    &0&.59927(8)  &0&.38(2)  &11&.35(2)  & 2&.114(7) & 2&.137(10)
&0&.989(2)  & 0&.993(3)\\\hline
0&.3025&21 &30    &0&.56525(19) &1&.02(5)$^*$&17&.02(5)   & 2&.786(10)&
2&.813(12) &0&.9903(8) & 0&.997(2)\\
0&.3025&30 &30    &0&.58479(10) &0&.79(5)$^*$&13&.24(3)   & 2&.338(7) &
2&.362(9)  &0&.990(2)  & 0&.999(2)\\\hline
0&.304 &30 &24    &0&.5827(4)   &2&.6(3)$^*$ &17&.52(9)   & 2&.839(8) &
2&.866(11) &0&.9907(13)& 0&.998(2)\\\hline
0&.305& 15 &36    &0&.53415(8)  &0&.52(2)    &26&.86(7)   & 3&.782(16)&
3&.82(3)   &0&.989(3)  & 0&.996(4)\\
0&.305& 21 &30    &0&.54098(10) &0&.73(3)  &24&.14(6)   & 3&.523(10)& 3&.57(2)
 &0&.988(2)  & 0&.998(3)\\
0&.305& 30 &30    &0&.54658(14) &1&.05(10)$^*$&22&.13(5)   & 3&.320(12)&
3&.35(2)   &0&.990(2)  & 1&.000(3)\\\hline
0&.31 &  6 &60    &0&.48187(2)  &0&.257(3) &65&.2(2)    & 6&.63(3)  & 6&.69(4)
 &0&.9909(10)& 1&.003(3)\\
0&.31 &  9 &48    &0&.50030(4)  &0&.302(6) &47&.25(12)  & 5&.44(3)  & 5&.49(5)
 &0&.9908(17)& 0&.998(5)\\
0&.31 & 15 &45    &0&.51178(4)  &0&.354(7) &39&.06(10)  & 4&.80(2)  & 4&.84(3)
 &0&.9911(12)& 1&.001(2)\\
0&.31 & 21 &42    &0&.51548(6)  &0&.41(2)  &36&.66(12)  & 4&.61(2)  & 4&.65(3)
 &0&.9915(10)& 0&.999(3)\\\hline
0&.32 &  9 &66    &0&.47234(3)  &0&.252(3) &82&.4(3)    & 7&.70(4)  & 7&.78(7)
 &0&.990(2)  & 1&.002(3)\\
0&.32 & 15 &66    &0&.48072(2)  &0&.264(5) &70&.5(3)    & 6&.96(4)  & 7&.04(4)
 &0&.9892(11)& 1&.000(3)\\\hline
0&.33 &  6 &102   &0&.43706(2)  &0&.212(3) &175&.3(2.0) &12&.13(16) &12&.23(21)
 &0&.991(3)  & 1&.001(9)\\
\end{tabular}
\end{table}

\begin{table}
\squeezetable
\caption{Strong coupling results at $N=\infty$.}
\label{SCtable}
\begin{tabular}{r@{}lr@{}lr@{}lr@{}lr@{}lr@{}lr@{}l}
\multicolumn{2}{c}{$\beta$}&
\multicolumn{2}{c}{$E$}&
\multicolumn{2}{c}{$C$}&
\multicolumn{2}{c}{$\chi$}&
\multicolumn{2}{c}{$\xi_G$}&
\multicolumn{2}{c}{$\xi_w$} \\
\tableline
0&.20 &0&.7814220 & 0&.052534 & 2&.93042 & 0&.78783 & 0&.83547 \\
0&.25 &0&.7090132 & 0&.101253 & 4&.53371 & 1&.14122 & 1&.17106 \\
0&.28 &0&.6556834 & 0&.154155 & 6&.24293 & 1&.52257 & 1&.51843 \\
0&.29 &0&.6352185 & 0&.179292 & 7&.01904 & 1&.71989 & 1&.68452 \\
0&.30 &0&.6129234 & 0&.210053 & 7&.93314 & 1&.99083 & 1&.89275 \\
\end{tabular}
\end{table}

\end{document}